\documentclass[3p]{elsarticle}
\usepackage{color}
\usepackage{amssymb}
\usepackage{graphicx}
\usepackage{amsmath}
\usepackage{subfig}
\usepackage{booktabs}
\usepackage{multirow}
\usepackage{afterpage}
\usepackage{tikz}
\usepackage{pgfplots}
\usepackage{pdflscape}
\usepgfplotslibrary{external}
\tikzset{external/system call={lualatex \tikzexternalcheckshellescape -halt-on-error -interaction=batchmode -jobname "\image" "\texsource"}}
\usepackage{placeins}
\usepackage{morefloats}

\journal{Ocean Engineering}

\begin{document}

\begin{frontmatter}

\title{Time-frequency analysis of ship wave patterns in shallow water:\\ modelling and experiments}

\author[Scott]{Ravindra Pethiyagoda}
\author[Scott]{Timothy J. Moroney}
\author[Gregor]{\\Gregor J. Macfarlane}
\author[Gregor]{Jonathan R. Binns}
\author[Scott]{Scott W. McCue\corref{cor1}}
\ead{scott.mccue@qut.edu.au}
%\ead[url]{http://www.elsevier.com}

\cortext[cor1]{Corresponding author: Tel: +61 (0)7
31384295, Fax: +61 (0)7 31381508}

\address[Scott]{School of Mathematical Sciences, Queensland University of Technology, QLD 4001, Australia}
\address[Gregor]{Australian Maritime College, University of Tasmania, Launceston, TAS 7248, Australia}

\begin{abstract}

A spectrogram of a ship wake is a heat map that visualises the time-dependent frequency spectrum of surface height measurements taken at a single point as the ship travels by.  Spectrograms are easy to compute and, if properly interpreted, have the potential to provide crucial information about various properties of the ship in question.  Here we use geometrical arguments and analysis of an idealised mathematical model to identify features of spectrograms, concentrating on the effects of a finite-depth channel.  Our results depend heavily on whether the flow regime is subcritical or supercritical.  To support our theoretical predictions, we compare with data taken from experiments { we} conducted in a model test basin using a variety of realistic ship hulls.  Finally, we note that vessels with a high aspect ratio appear to produce spectrogram data that contains periodic patterns.  We can reproduce this behaviour in our mathematical model by using a so-called two-point wavemaker. {These results highlight the role of wave interference effects in spectrograms of ship wakes.}

\end{abstract}


\end{frontmatter}
\section{Introduction}

There is recent interest in applying time--frequency analysis as a tool for capturing various features of a ship wake.  Such analysis involves performing many short-time Fourier transforms on a cross-section of a ship wake and visualising the resulting data as a spectrogram \cite{brown89,didenkulova13,sheremet13,torsvik15a,torsvik15b,wyatt88}.  As this approach only requires measurements taken at a single stationary sensor as a ship sails past \cite{brown89,david17,parnell08}, it offers a  viable method for analysing ship waves in real-world conditions.  This line of research has a range of potential practical applications in terms of quantifying the negative effects that a propagating wake wash will have when it interacts with a coastal zone \cite{torsvik15a} or facilitating remote sensing and surveillance of unmonitored vessels.

Before now, the published studies on spectrogram analysis of ship wakes use experimental measurements taken from an open body of water \cite{brown89,parnell08,sheremet13}. As such, the surface height data involves a complicated combination of information from multiple ships and wind waves.  The resulting interference has made it difficult for previous researchers to confidently attribute features of a spectrogram to various properties of a ship or its wake \cite{torsvik15a}.  In response to this challenge, Pethiyagoda et al.~\cite{pethiyagoda17} used linear water wave theory together with a  mathematical model to explain features of spectrograms of ship waves that are small in amplitude.  This explanation includes a derivation of the linear dispersion curve which predicts the location of the colour intensity on the spectrogram when the ship in question is travelling through an infinitely deep body of water.  Further, they used a simple weakly nonlinear theory and simulations of the fully nonlinear version of the model to identify features of a spectrogram that are due to nonlinearity.  The comparison between these new theoretical results with data measured in a shipping channel in the  Gulf of Finland is promising \cite{pethiyagoda17}; however, there is a need to extend the analysis to hold for finite-depth channels and to test the predictions against cleaner data gathered from controlled experiments.

In this paper, we extend the analysis in \cite{pethiyagoda17} to apply to finite-depth channels.  Ship wakes on a finite-depth fluid are interesting because they can be classified as either subcritical ($F_H<1$, where $F_H$ is the depth based Froude number) or supercritical ($F_H>1$), with qualitatively different wave patterns forming for each case.  For example, wakes for subcritical flows are made up of transverse waves and divergent waves, while wakes for supercritical flows contain divergent waves only.  It is interesting that the wavelength of supercritical waves can vary greatly, and often contain some very long-wavelength waves which are potentially very damaging to shorelines in sheltered waterways, where such energetic waves do not occur naturally~\cite{macfarlane14}.
The difference between subcritical and supercritical flows has a noticeable effect on the linear dispersion relation which, as we show, directly affects the location of the high intensity regions on the spectrograms.  We support our theoretical findings by comparing with experimental surface height measurements {we have} taken from a model test basin, eliminating the effect of environmental factors such as wind waves, currents or varying bathymetry.

Our study begins in Section~\ref{sec:mathematicalmodel} by presenting our mathematical model, which involves inviscid fluid flow past an applied Gaussian-type pressure on the surface. Here the pressure distribution is used as an idealised mathematical representation of a ship and the strength of the pressure ($\epsilon$) acts as a proxy for the volume of water displaced by the ship.  In Section~\ref{sec:spectrograms} we extend the geometric argument for determining the linear dispersion curve presented by Pethiyagoda et al.~\cite{pethiyagoda17} to include the effects of a finite depth fluid. We observe the effect of changing the width of the Gaussian pressure relative to the water depth ($\delta$), where the regions of high intensity favour higher frequencies along the dispersion curves for smaller pressure widths.  Then, in Section~\ref{sec:experimental} we validate the new linear and second-order dispersion curves against spectrograms generated from experimental data {we} collected from the model test basin at the Australian Maritime College. We observe trends in the experimental spectrograms for different sailing speeds and hull shapes, and provide possible explanations for these trends in terms of linear and nonlinear wave properties.  Lastly, in Section~\ref{sec:interference}, certain periodic patterns in the spectrograms for vessels with a high aspect ratio are explained by taking into account wave interference effects.

\section{Mathematical model}\label{sec:mathematicalmodel}

\subsection{Problem setup}\label{sec:problemsetup}
In order to simulate a wake left behind a moving ship, we consider the idealised problem of calculating the free surface disturbance created by a steadily moving pressure distribution applied to the surface of a body of water of constant depth $H$.  We suppose the pressure distribution is of a Gaussian type with strength $P_0$ and characteristic width $L$, and then formulate the mathematical problem in the reference frame of this moving pressure.  We nondimensionalise the problem by scaling all velocities by the speed of the pressure distribution, $U$, and all lengths by $U^2/g$, where $g$ is acceleration due to gravity. The fully nonlinear governing equations are then
\begin{align}
\nabla^2\phi&=0& &\text{for }-F_H^{-2}<z<\zeta(x,y),\label{eq:laplace}\\
\label{eq:bern}
\frac{1}{2}|\nabla\phi|^2+\zeta+\epsilon p&=\frac{1}{2}& &\text{on }z=\zeta(x,y),\\
\phi_x\zeta_x+\phi_y\zeta_y&=\phi_z& &\text{on }z=\zeta(x,y),\\
\phi_z&=0& &\text{on }z=-F_H^{-2}\\
\phi&\sim x& &\text{as }x\rightarrow -\infty,\label{eq:upstream}
\end{align}
where $\phi(x,y,z)$ is the velocity potential, $\zeta(x,y)$ is the free-surface height, $\epsilon=P_0/(\rho U^2)$ is the dimensionless pressure strength, $\rho$ is the fluid density and $\epsilon p(x,y)$ is the pressure distribution. For the present study we will use the pressure distribution
\begin{equation}
p(x,y)=\mathrm{e}^{-\pi^2F_L^4(x^2+y^2)},\label{eq:pressure}
\end{equation}
where $F_H=U/\sqrt{gH}$ is the depth-based Froude number and $F_L=U/\sqrt{gL}$ is the length-based Froude number. We can also define the scaled width of the pressure distribution $\delta=L/H=F_H^2/F_L^2$.  In this formulation, $F_H$ is the parameter that measures the speed of the moving pressure, while the pressure strength $\epsilon$ provides a measure of nonlinearity in the problem (the regime $\epsilon\ll 1$ is approximately linear).

Our use of the applied pressure (\ref{eq:pressure}) to act as a mathematical representation of ship is deficient in the sense that it does not allow for precise features of the ship hull to be described in any way.  Despite this modelling simplification, it is common to use this idea when analysing ship wakes from a mathematical perspective \cite{darmon14,ellingsen14,li16,pethiyagoda15,smeltzer17}, and indeed we show that there are many advantages in this approach.  To understand the analogy with real ship vessels, it is worth interpreting $\epsilon$ as a crude measure of the volume of water displaced by a vessel.  Further, note that we can easily extend this model to include multiple pressure distributions, as we do later in Section~\ref{sec:interference}.

\subsection{Exact solution to linear problem}

For moderate to large values of $\epsilon$, the mathematical problem (\ref{eq:laplace})-(\ref{eq:pressure}) is highly nonlinear.  From a computational perspective, obtaining accurate numerical solutions is challenging as the problem is three-dimensional and the upper surface $z=\zeta(x,y)$ is unknown and must be solved for as part of the solution process.  Progress with the numerical solution to this problem and related problems can made using boundary integral methods \cite{forbes89,pethiyagoda14b,pethiyagoda14a,parau02}, for example.

On the other hand, for weak pressure distributions, $\epsilon\ll 1$, the problem (\ref{eq:laplace})-(\ref{eq:pressure}) can be linearised.  The linearised version has the exact solution \cite{wehausen60}
\begin{align}
\zeta(x,y) = &-\epsilon p(x,y)+\frac{\epsilon}{2\pi^2}\int\limits_{-\pi/2}^{\pi/2}
\,\,\int\limits_{0}^{\infty}\frac{k^2\tilde{p}(k,\theta)\cos(k[|x|\cos\theta+y\sin\theta])}{k-\sec^2\theta\tanh (k/F_H^2)}\,\,\mathrm{d}k\,\mathrm{d}\theta\notag
\\
&-\frac{2\epsilon F_H^2 H(x)}{\pi}\int\limits_{\theta_0}^{\pi/2}\frac{k_0^2\tilde{p}(k_0,\theta) \sin(k_0x\cos\theta)\cos(k_0y\sin\theta)}{F_H^2-\sec^2\theta\,\mathrm{sech}^2(k_0/F_H^2)}\,\mathrm{d}\theta,\label{eq:exactLinearFinite}
\end{align}
where $\theta_0=0$ for $F_H<1$ and $\theta_0=\arccos(1/F_H)$ for $F_H>1$,
$$\tilde{p}(k,\theta)=\exp(-k^2/(4\pi^2F_L^4))/(\pi F_L^4)$$ is the Fourier transform of the pressure distribution (\ref{eq:pressure}), $H(x)$ is the Heaviside function and the path of integration over $k$ is taken below the pole $k=k_0(\theta)$, where $k_0(\theta)$ is the real positive root of
\begin{equation}
k-\sec^2\theta\tanh (k/F_H^2)=0,\qquad \theta_0<\theta<\frac{\pi}{2}.
\label{dispersion1}
\end{equation}
Note (\ref{dispersion1}) is the linear dispersion relation for steady ship wave patterns which we discuss in some detail in Section~\ref{sec:dispCurve}.

\begin{figure}
\centering
\subfloat[Subcritical ($F_H=0.6$)]{\includegraphics[width=.5\linewidth]{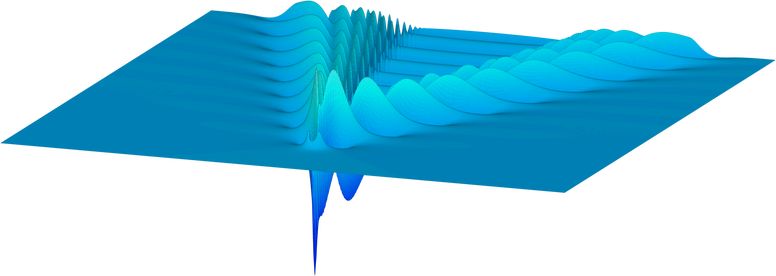}}\hfil
\subfloat[Supercritical ($F_H=1.34$)]{\includegraphics[width=.5\linewidth]{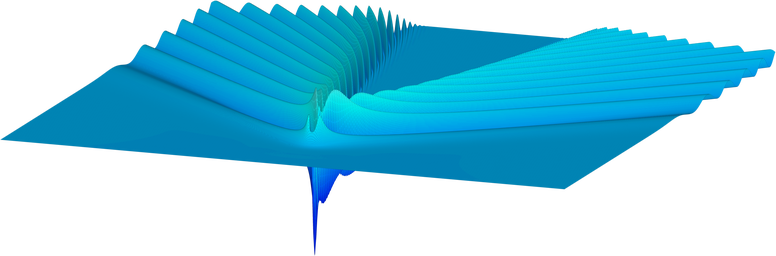}}
\caption{Free-surface profiles for linear flow past the pressure distribution (\ref{eq:pressure}) in a fluid of finite depth for (a) subcritical ($F_H=0.6$) and (b) supercritical ($F_H=1.34$) flows where $F_L=1$.}
\label{fig:LinSurface}
\end{figure}

Figure \ref{fig:LinSurface} presents free-surface profiles calculated using the exact linear solution (\ref{eq:exactLinearFinite}) for a subcritical Froude number, $F_H=0.6$, and a supercritical Froude number, $F_H=1.34$.  In the subcritical case we see that the wave pattern is comprised of transverse waves that run perpendicular to the direction of travel and divergent waves that are oblique to this direction.  On the other hand, as noted in the Introduction, the wave pattern for the supercritical case contains only divergent waves.

\section{Spectrograms of finite-depth ship wakes}\label{sec:spectrograms}

\subsection{Computing a spectrogram}
To compute the spectrogram data for a given signal, $s(t)$, we take the square magnitude of a short-time Fourier transform
\begin{equation}
S(t,\omega)=\left|\int\limits_{-\infty}^{\infty}h(\tau-t)s(\tau)\mathrm{e}^{-i\omega\tau}\,\mathrm{d}\tau\right|^2,\label{eq:spec}
\end{equation}
where $h(t)$ is a window function which is  even with compact support. In this paper we will use the Blackman-Harris 92dB window function \cite{harris78}. The results are presented in a time-frequency heat map of dimensionless angular frequency $\omega$ against scaled time $t/y$ (with $y$ held constant) with the colour intensity aligned to $\log_{10}(S(t,\omega))$.

The signal $s(t)$ comes from fixing $y$ in the solution to be $y=y_s$, say, and taking $s(t)=\zeta(t,y_s)$, where we recall the nondimensional speed is unity ($x=t$).  For the spectrograms generated by the exact solution to the linear problem (\ref{eq:exactLinearFinite}), we have chosen to fix $y_s=100$, but it is important to note that for sufficiently large $y_s$, the spectrograms are visually extremely similar.

\subsection{Linear dispersion curve}\label{sec:dispCurve}
\begin{figure}
\centering
\subfloat[Subcritical ($F_H=0.6$)]{\includegraphics[width=.45\linewidth]{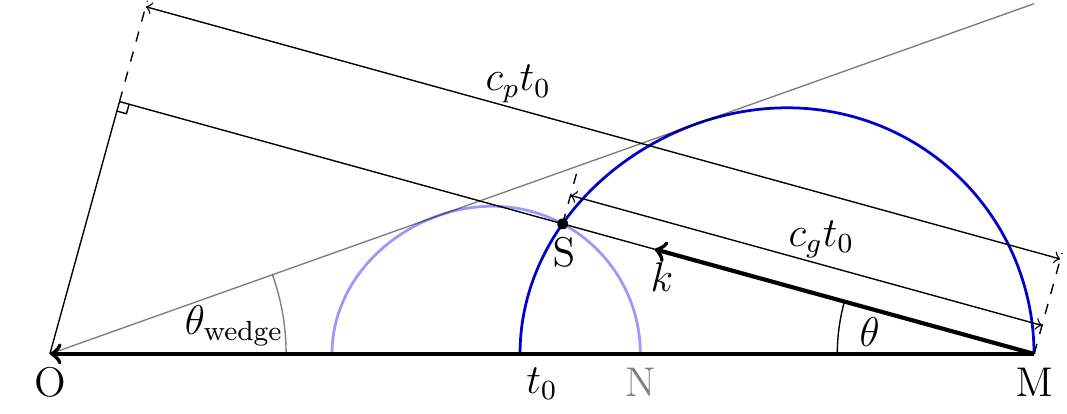}}\hfil
\subfloat[Supercritical ($F_H=1.34$)]{\includegraphics[width=.45\linewidth]{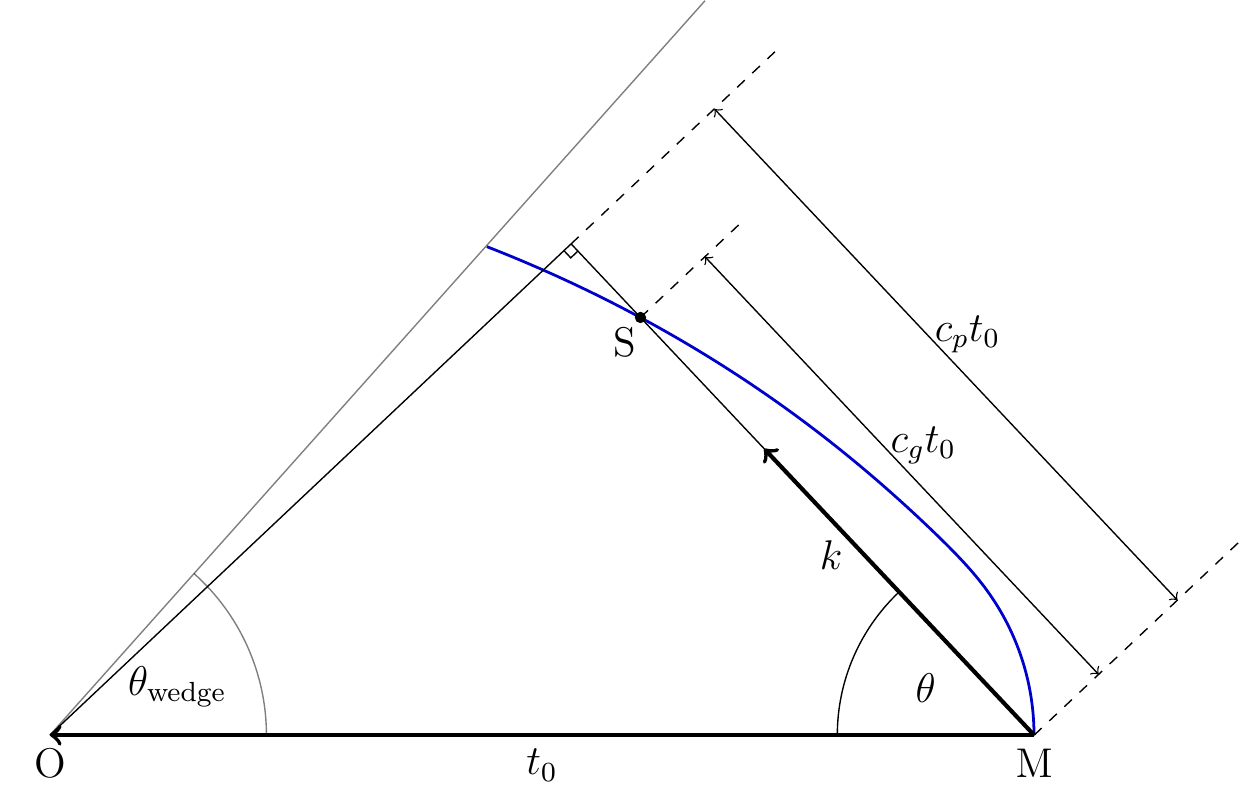}}
\caption{These schematics of ship wave dispersion for (a) subcritical ($F_H=0.6$) and (b) supercritical ($F_H=1.34$) flows illustrates a ship moving from the point M to the point O with a nondimensional speed 1 over the time period $t_0$. The waves generated by the ship at the point M propagate at angle $\theta$  from the sailing line, and wavenumbers $k$ with phase velocity, $c_p$,  and group velocity, $c_g$. The group velocity curve (dark blue) originating from the point M is traced out using the group velocity for angle $\theta$ where (a) $0\leq\theta\leq\pi/2$ or (b) $\arccos(1/F_H)\leq\theta\leq\pi/2$. The light blue curve originating from point N in (a) is another group velocity curve that follows the same rules as the dark blue curve but is generated at a later time. The grey line, represents the geometric edge of the wake system with a wedge angle $\theta_\mathrm{wedge}$. Finally, the point S is an arbitrary point along the dark blue group velocity curve.}
\label{fig:SpecDiagram}
\end{figure}

By following Pethiyagoda et al.~\cite{pethiyagoda17} we now use geometric arguments to determine the linear dispersion curve for ship waves in a finite depth channel.  The key ideas here are borrowed from standard texts \cite{lamb16,lighthill78,wehausen60} {and articles} \cite{havelock08}.  First we note that, monochromatic finite-depth linear water waves obey the dispersion relation $\omega=\Omega(k)$, where $\omega$ is the wave frequency in the reference frame of the observer,
\begin{equation}
\Omega(k)=\sqrt{k\tanh\left(\frac{k}{F_H^2}\right)}\label{eq:dispersion}
\end{equation}
is the wave function, and $k$ is the wavenumber.   This result implies that the phase and group velocities are given by
\begin{align}
c_p&=\frac{\Omega(k)}{k}=\frac{\sqrt{\tanh\left(\frac{k}{F_H^2}\right)}}{\sqrt{k}},\label{eq:phaseVelgen}\\ c_g&=\frac{\mathrm{d}\Omega}{\mathrm{d}k}=\frac{\sqrt{\tanh\left(\frac{k}{F_H^2}\right)}}{2\sqrt{k}}\left(1+\frac{2k}{F_H^2\sinh\left(\frac{2k}{F_H^2}\right)}\right).\label{eq:groupVel}
\end{align}
For the following we also define the ratio of group velocity to phase velocity
\begin{equation}
\alpha(k)=\frac{c_g}{c_p}=\frac{1}{2}\left(1+\frac{2k}{F_H^2\sinh\left(\frac{2k}{F_H^2}\right)}\right)\label{eq:alphaFunc}
\end{equation}
At this point it is worth emphasising that the wave function (\ref{eq:dispersion}) and the wave velocities (\ref{eq:phaseVelgen})-(\ref{eq:groupVel}) depend on the Froude number $F_H$, whereas in the infinite depth limit, there is no dependence on velocity (as $F_H\rightarrow 0$, $\Omega\sim k^{1/2}$, $c_p\sim k^{-1/2}$, $c_g\sim k^{-1/2}/2$, $\alpha\rightarrow 1/2$).

In order to compute the linear dispersion curve, we consider a ship moving along the path MO a distance of $t_0$ as shown in Figure \ref{fig:SpecDiagram}, where we recall the ship's dimensionless speed is unity.  At the point M the ship will generate waves at an angle $\theta$ from MO.  In the reference frame of the ship, the doppler-shifted dispersion relation becomes $(\omega_{\mbox{ship}}-k\cos\theta)^2=\Omega^2(k)$.  In the special case of a stationary wave pattern relative to the ship, the frequencies $\omega_{\mbox{ship}}$ all vanish, which gives
\begin{equation}
\Omega(k)=k\cos\theta.
\label{dispersion2}
\end{equation}
The steady dispersion relationship (\ref{dispersion2}), together with (\ref{eq:dispersion}), provides an implicit relationship between $k$ and $\theta$. Note that (\ref{dispersion2}) is equivalent to (\ref{dispersion1}).   Thus there are bounds on the possible directions that $\theta$ may take.  For subcritical flows ($F_H<1$), $c_p\leq 1$ for $k\geq k_0(0)$, where $k_0(0)$ is the positive root of (\ref{dispersion1}) with $\theta=0$.  Thus, $\theta$ takes all the angles from $\theta=0$ (when $k=k_0(0)$) to $\theta=\pi/2$ (as $k\rightarrow\infty$).  For supercritical flows ($F_H>1$), $c_p\rightarrow 1/F_H<1$ as $k\rightarrow 0$ and so $\theta$ lies in the interval between $\theta=\arccos(1/F_H)$ (when $k=0$) to $\theta=\pi/2$ (as $k\rightarrow\infty$).  Further, note we can rewrite (\ref{dispersion2}) as
\begin{equation}
c_p=\cos\theta,\label{eq:phaseVel}
\end{equation}
which we shall use in the following working.

Referring again to Figure \ref{fig:SpecDiagram}, only certain waves produced at M can contribute to a stationary ship wave pattern.  The energy associated with these waves travels from M at the group velocity to points on the dark blue curve in Figure \ref{fig:SpecDiagram} over the time period $t_0$.  We refer to this curve as the group velocity curve.  The $x$ and $y$ coordinates of points on this curve, $x_s$ and $y_s$, are given by $x_s= t_0-c_gt_0\cos\theta$, $y_s= c_gt_0\sin\theta$.  By substituting in (\ref{eq:alphaFunc}) and (\ref{eq:phaseVel}), we arrive at the parametric representation
\begin{align}
x_s &=t_0-\alpha(k)t_0\cos^2\theta,\label{eq:sensorX}\\
y_s &=\alpha(k)t_0\cos\theta\sin\theta,\label{eq:sensorY}
\end{align}
where we emphasise that $k$ is a function of $\theta$ via (\ref{dispersion2}).  Equations (\ref{eq:sensorX})--(\ref{eq:sensorY}) can be combined to form the equation
\begin{equation}
0=\tan^2\theta-\frac{x_s}{y_s}\alpha(k)\tan\theta+1-\alpha(k).\label{eq:thetaZero}
\end{equation}
Thus we have an expressions that relates the direction of a wave leaving M ($\theta$) with a wavenumber ($k$) and a position relative to the ship when the ship is at O.

We note in passing that, with this picture, we can compute the wedge angle $\theta_\mathrm{wedge}$ that encloses
all wave generation
by observing that there is a maximum angle at which rays can leave O and intersect the group velocity curve (dark blue).  That is,  $\theta_\mathrm{wedge}=\max\lbrace\arctan(y_s/x_s)\rbrace$.  For the infinite depth case $F_H=0$, equation (\ref{eq:thetaZero}) reduces to a simple quadratic (since $\alpha=1/2$ in that case) which gives the well-known Kelvin angle $\theta_\mathrm{wedge}=\arctan(1/\sqrt{8})$.  For subcritical flows, the calculation for $\theta_\mathrm{wedge}$ must be performed numerically.  In the supercritical case, the maximum occurs when $\theta=\arccos(1/F_H)$ (see Figure \ref{fig:SpecDiagram}(b)), which gives the result $\theta_\mathrm{wedge}=\arctan(1/\sqrt{F_H^2-1})$.

\begin{figure}[htb]
	\centering
	\includegraphics[width=.5\linewidth]{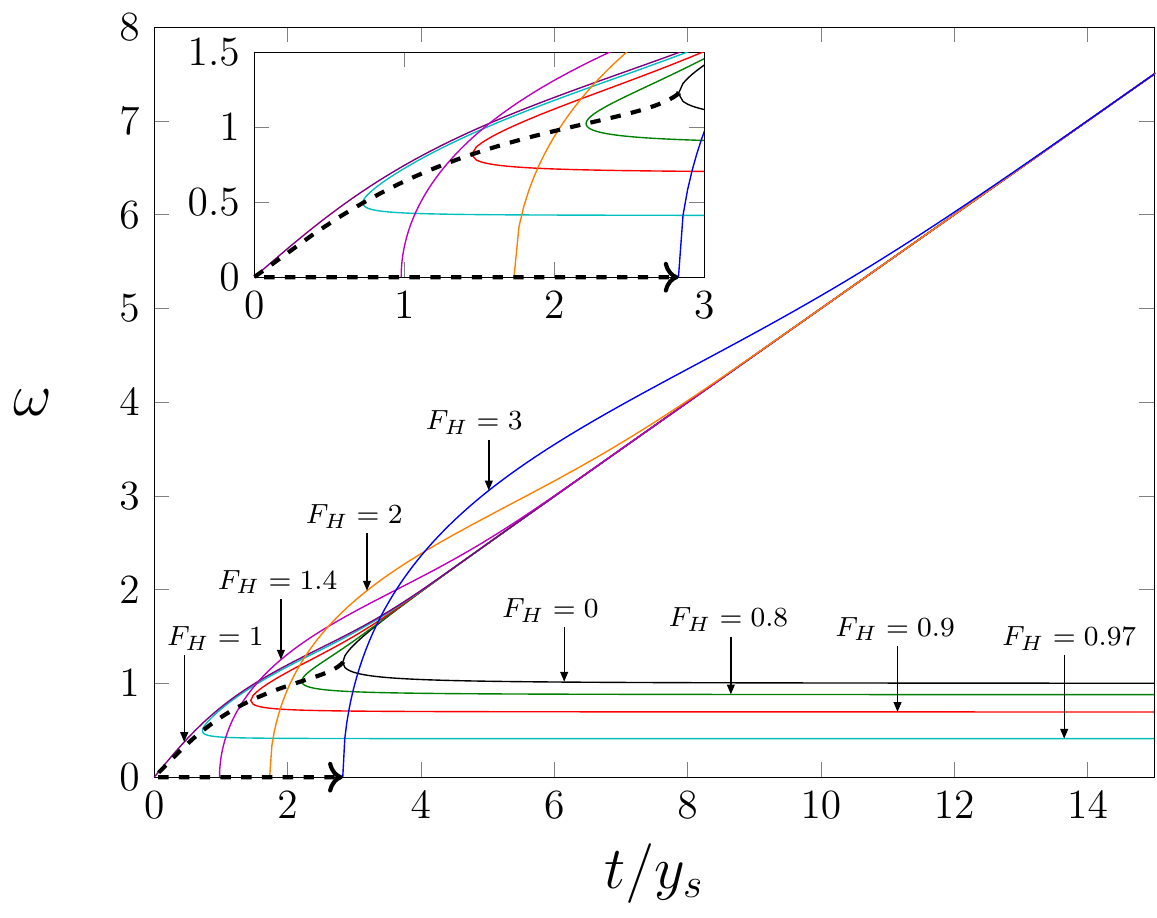}
	\caption{The linear dispersion curve described in Section \ref{sec:dispCurve} is plotted for {$F_H=0$, $0.8$, $0.9$, $0.97$, $1$, $1.4$, $2$, and $3$}. The dashed curve follows the earliest point of the dispersion curve with the arrow indicating the direction of increasing $F_H$. { Note, the $F_H=1$ curve touches the horizontal axis at $t/y_s=0$.} An inset of the range $0<t/y_s<3$ is also provided.}
	\label{fig:LinDisp}
\end{figure}

If a sensor is at given position relative to the ship, $(x_s,y_s)$, then for subcritical flows there are two group velocity curves (a dark blue curve and a light blue curve) that pass through $(x_s,y_s)$ (see Figure~\ref{fig:SpecDiagram}(a)) and so two pairs  of $\theta$ and $k$ which we label $\theta_i$ and $k_i$ for $i=1,2$.
These are found by solving (\ref{eq:thetaZero}) numerically.  For supercritical flows there is only one relevant group velocity curve (dark blue) and so only one solution to (\ref{eq:thetaZero}) for $\theta$ and $k$ (which we call $\theta_1$ and $k_1$).  Either way, for a stationary sensor at $(x_s,y_s)$, the angular frequency observed by the sensor is described by $\omega=\Omega(k)$, where $\Omega(k)$ is given by (\ref{dispersion2}).  As there are possible options for $k$, we can write
\begin{equation}
\omega_i=k_i\cos\theta_i.
\label{eq:observedFrequency}
\end{equation}
Finally, to generate the linear dispersion curve, we plot $\omega_i$ versus $x_s/y_s$.  Note that when we draw our dispersion curves on a spectrogram, the sensor relative to the ship is moving with dimensionless speed unity, so our plots are actually of $\omega_i$ versus $t/y_s$.\footnote{While we are only considering stationary sensors in this study, we note that if the sensor is moving with dimensionless speed $U_s$ in the direction $\theta_s$ where $\theta_s$ is taken relative to the sailing line of the ship, the doppler-shifted dispersion relation gives
\begin{equation*}
\omega_i=k_i\cos\theta_i-U_sk_i\cos\left(\theta_i-\theta_s\right).\label{eq:observedFrequencyGeneral}
\end{equation*}
In this case, the dispersion curve is drawn by defining the location of the sensor as a function of time, $(x_s,y_s)=(x_0-t(1+U_s\cos\theta_s),y_0+tU_s\sin\theta_s)$, where $x_0$ and $y_0$ are the location of the sensor at $t=0$.}

A plot of the linear dispersion curve for a variety of depth-based Froude numbers is presented in Figure~\ref{fig:LinDisp}.  In each case, these curves help to predict the location of the high intensity regions of a spectrogram of a ship wake (see Section~\ref{sec:linearspectrogram}).  The special case $F_H=0$ is the linear dispersion curve for an infinitely deep fluid, as calculated in Pethiyagoda et al.~\cite{pethiyagoda17}.  It is made up of two branches.  The upper branch (corresponding to $\omega_1$) indicates a sliding frequency mode which represents the divergent waves, while the lower branch ($\omega_2$) describes a constant frequency mode representing the transverse waves.  The left-most point on the linear dispersion curve, $\min\lbrace t/y_s\rbrace=\sqrt{8}$, gives the first time that the wave reaches the sensor, again providing the Kelvin wake angle since the sensor relative to the ship is $(x_s,y_s)=(t,y_s)$.

Moving on to the finite-depth case, the sliding frequency mode of each dispersion curve behaves like $\omega\sim t/(2y_s)$ as $t/y_s\rightarrow \infty$.  This property is shared by all dispersion curves, regardless of Froude number.  For subcritical Froude numbers ($F_H<1$), the linear dispersion curve is qualitatively similar to the infinite depth case.  Quantitatively, we see that the frequency of the $\omega_2$ mode decreases as $F_H$ increases until the limit $\omega_2\rightarrow 0$ as $F_H\rightarrow 1^-$, which is consistent with the property that the wavelength of transverse waves increases to infinity in this critical limit.  The left-most point $\min\lbrace t/y_s\rbrace$ also decreases as $F_H$ increases, meaning the wave pattern reaches the sensor earlier, reflecting the result that the wedge angle increases for subcritical Froude numbers until $\theta_\mathrm{wedge}\rightarrow\pi/2^-$ as $F_H\rightarrow 1^-$.  For supercritical Froude numbers ($F_H>1$), the linear dispersion curve is qualitatively different, as there is now only one branch, for divergent waves ($\omega_1$ branch).  This curve intersects the $t/y_s$ axis at $t/y_s=\sqrt{F_H^2-1}$ and is monotonically increasing with $t/y_s$.

\subsection{Weakly nonlinear predictions}\label{sec:nonlineardispcurves}

In the infinite-depth case, Pethiyagoda et al.~\cite{pethiyagoda17} provide a simple formula for determining the second-order dispersion curves, which come from a weakly nonlinear approximation.  For subcritical flows, the idea carries over directly, to give $\omega_{ij\pm}=\omega_i\pm\omega_j$ for $i,j=1,2$, where we recall $\omega_1$ represents the divergent waves while $\omega_2$ represents the transverse waves.  (In our recent work \cite{pethiyagoda17}, we used the notation $\omega_3=\omega_{11+}$, $\omega_4=\omega_{22+}$, $\omega_5=\omega_{12+}$, $\omega_6=\omega_{12-}$.) For supercritical flows there are no transverse waves, so the second-order dispersion curve is simply $\omega_{11+}=2\omega_1$.  We return to these weakly nonlinear approximations in Section \ref{sec:experimental}.

\subsection{Linear spectrogram}\label{sec:linearspectrogram}

\begin{figure}
\begin{tabular}{ccc}
\hspace{.11\linewidth}$F_H=0.6$&$F_H=1.34$&\\[1ex]
\hspace{2ex}\raisebox{11ex}{\rotatebox{90}{$F_L=0.3$}}
\includegraphics[width=.395\linewidth]{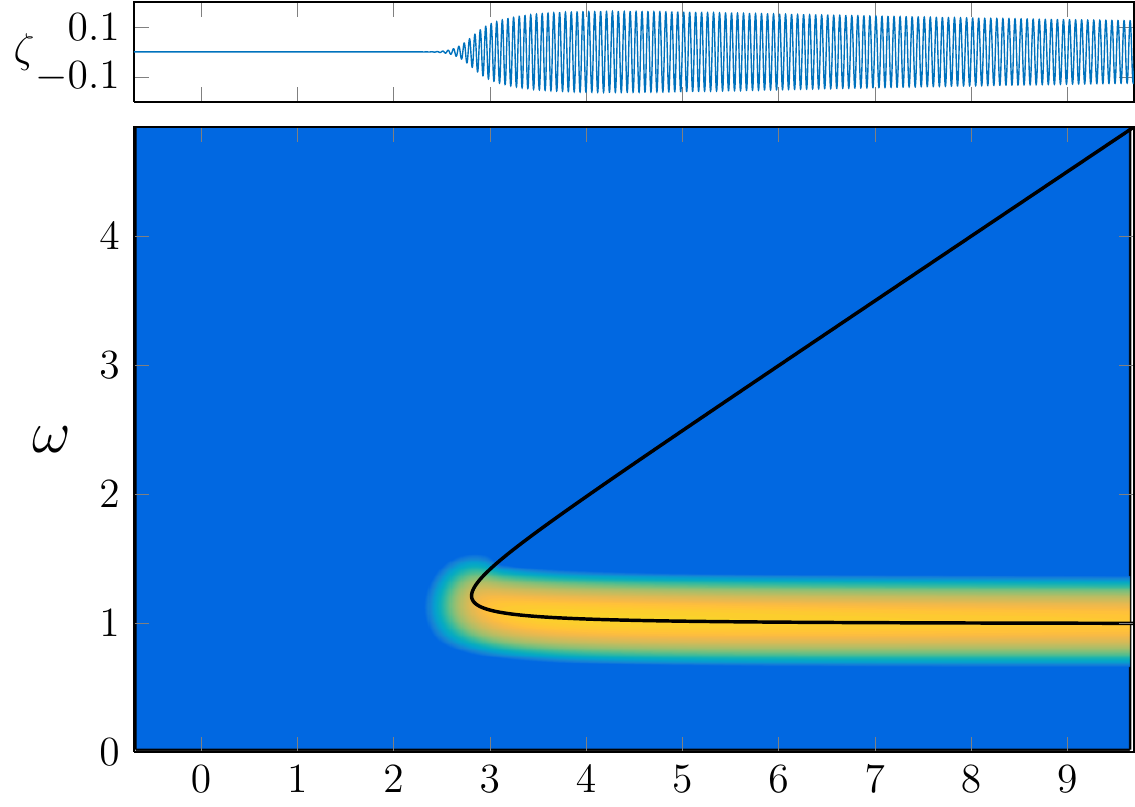}&
\includegraphics[width=.385\linewidth]{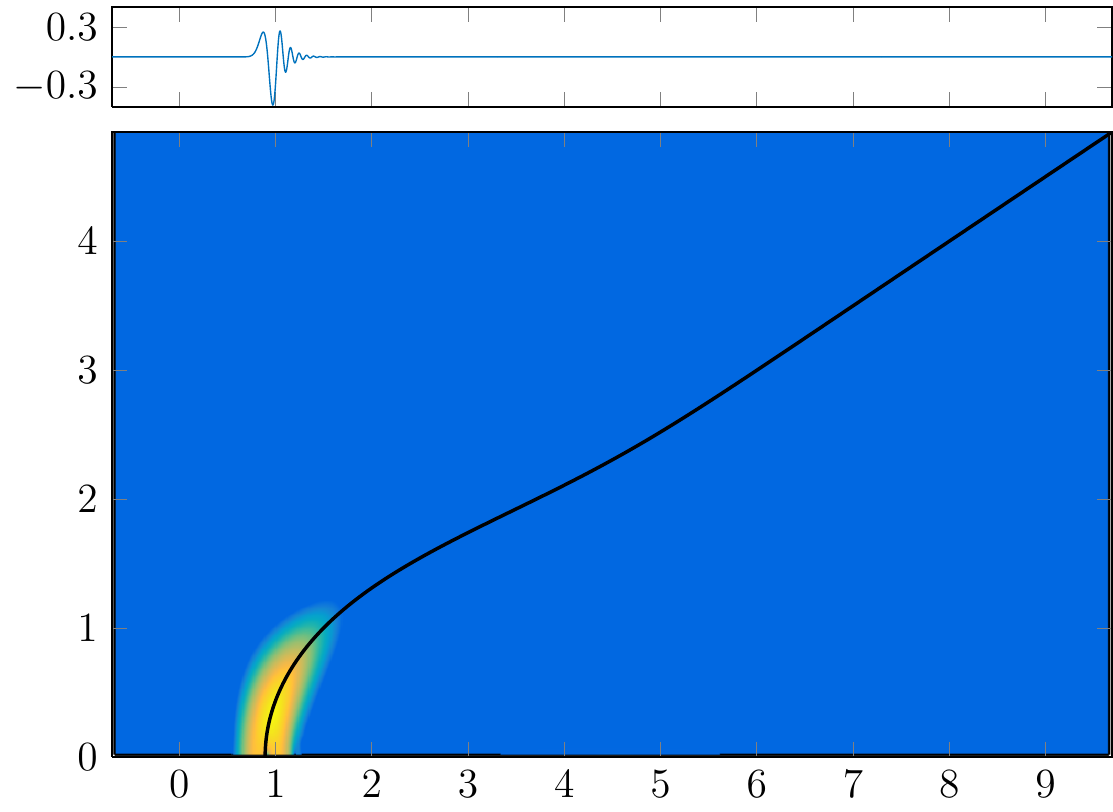}&\\
\hspace{2ex}\raisebox{11ex}{\rotatebox{90}{$F_L=0.7$}}
\includegraphics[width=.395\linewidth]{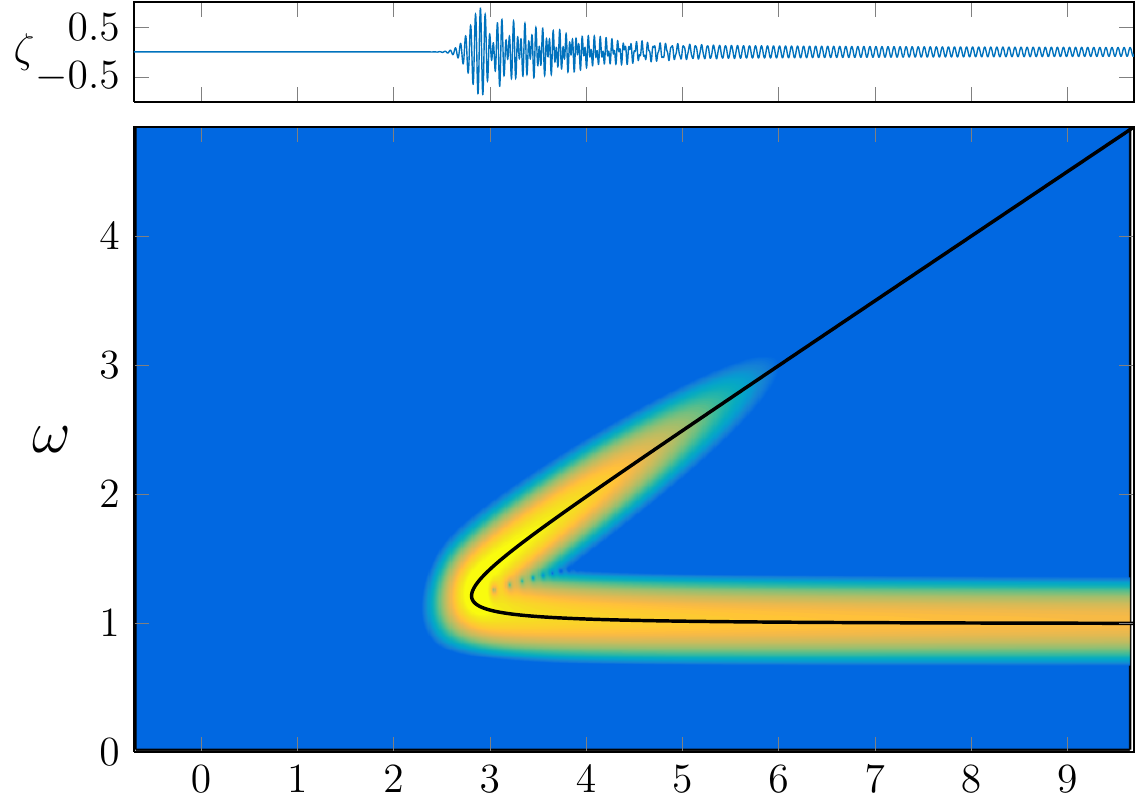}&
\includegraphics[width=.385\linewidth]{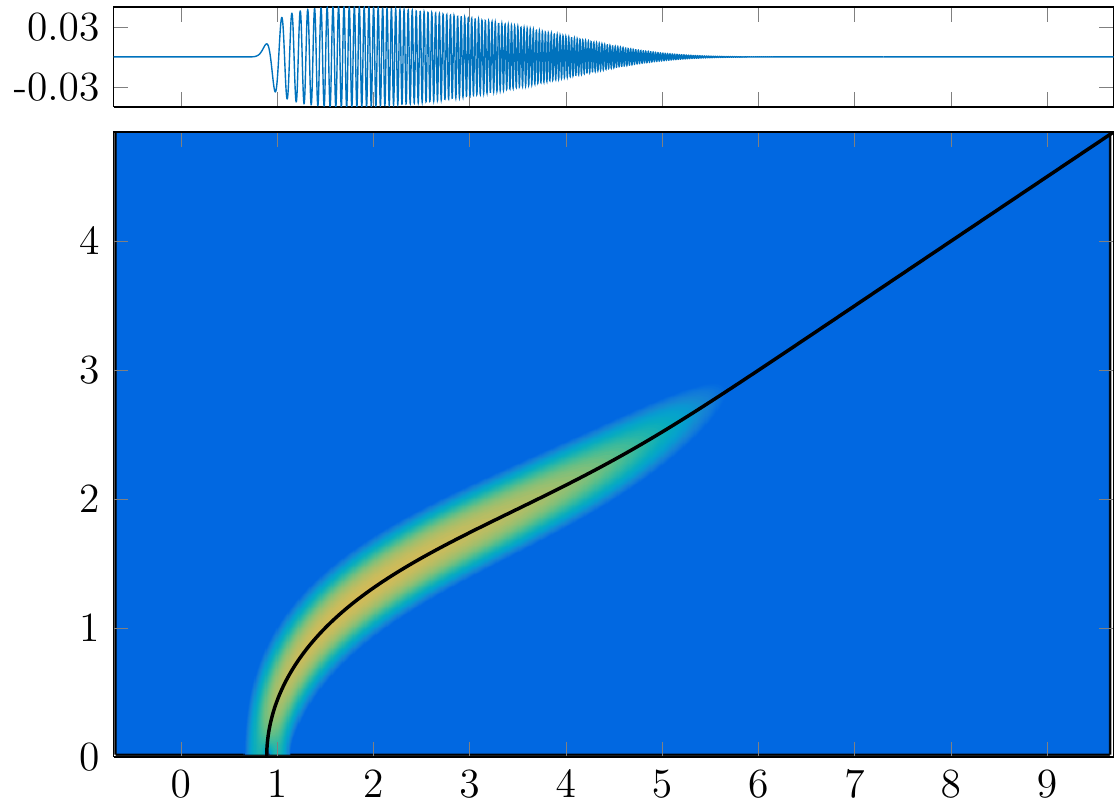}&
\hspace{-.02\linewidth}\multirow{4}{*}[0.17\linewidth]{\includegraphics[width=.06\linewidth]{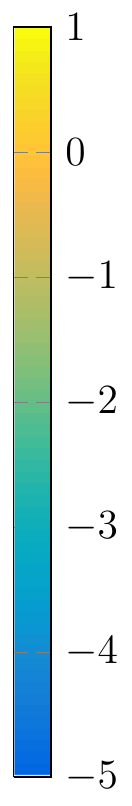}}\\
\hspace{2ex}\raisebox{11ex}{\rotatebox{90}{$F_L=1$}}
\includegraphics[width=.395\linewidth]{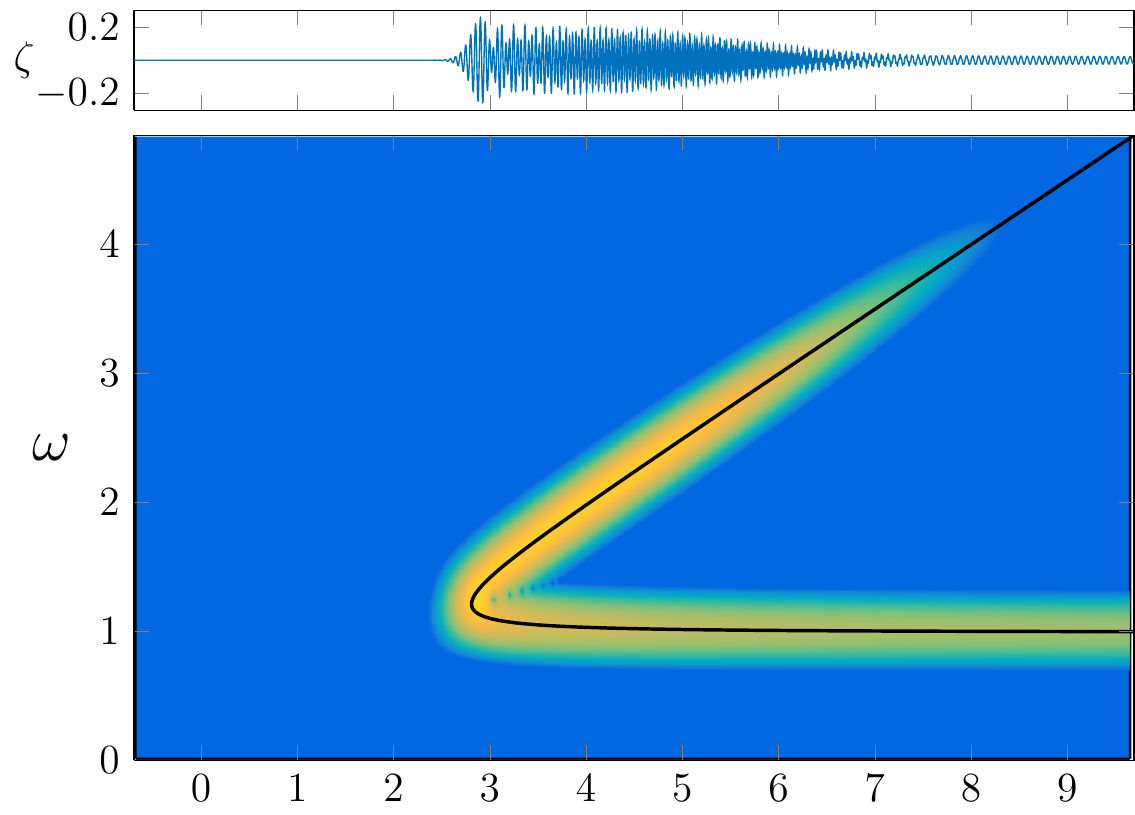}&
\includegraphics[width=.385\linewidth]{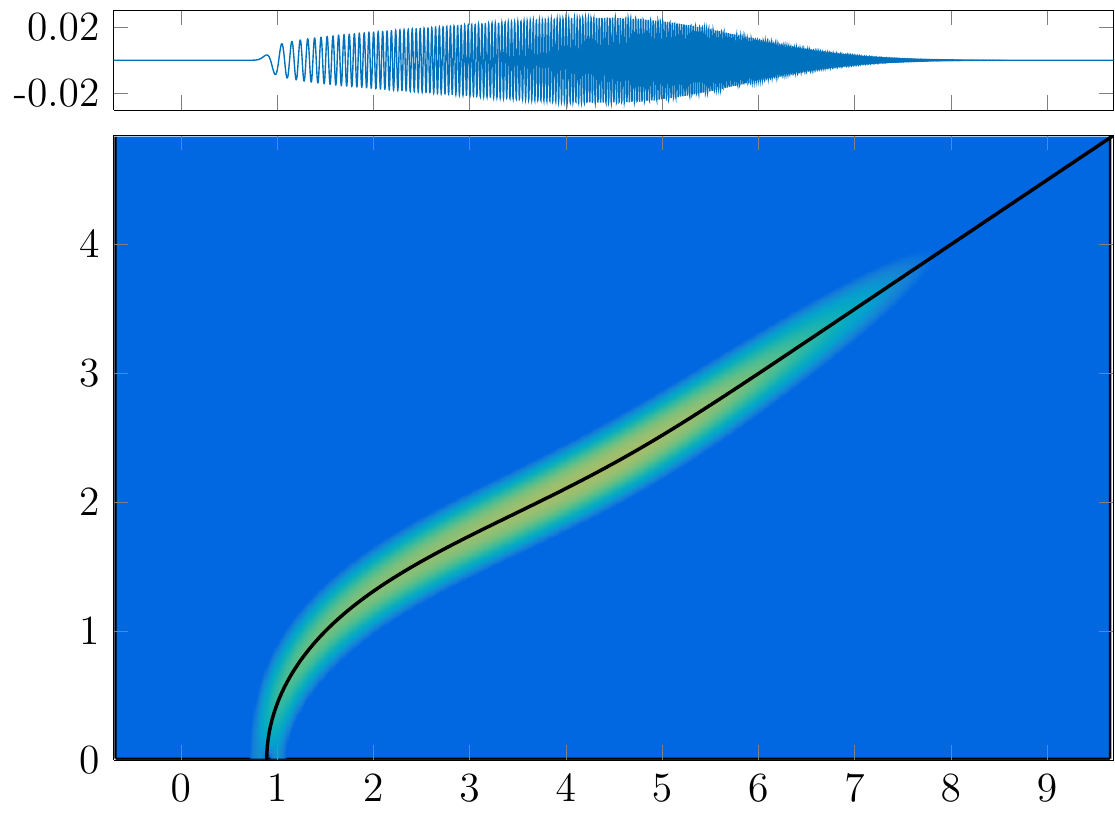}&\\[-1ex]
\hspace{2ex}\raisebox{14ex}{\rotatebox{90}{$F_L=1.5$}}
\subfloat[Subcritical]{\includegraphics[width=.395\linewidth]{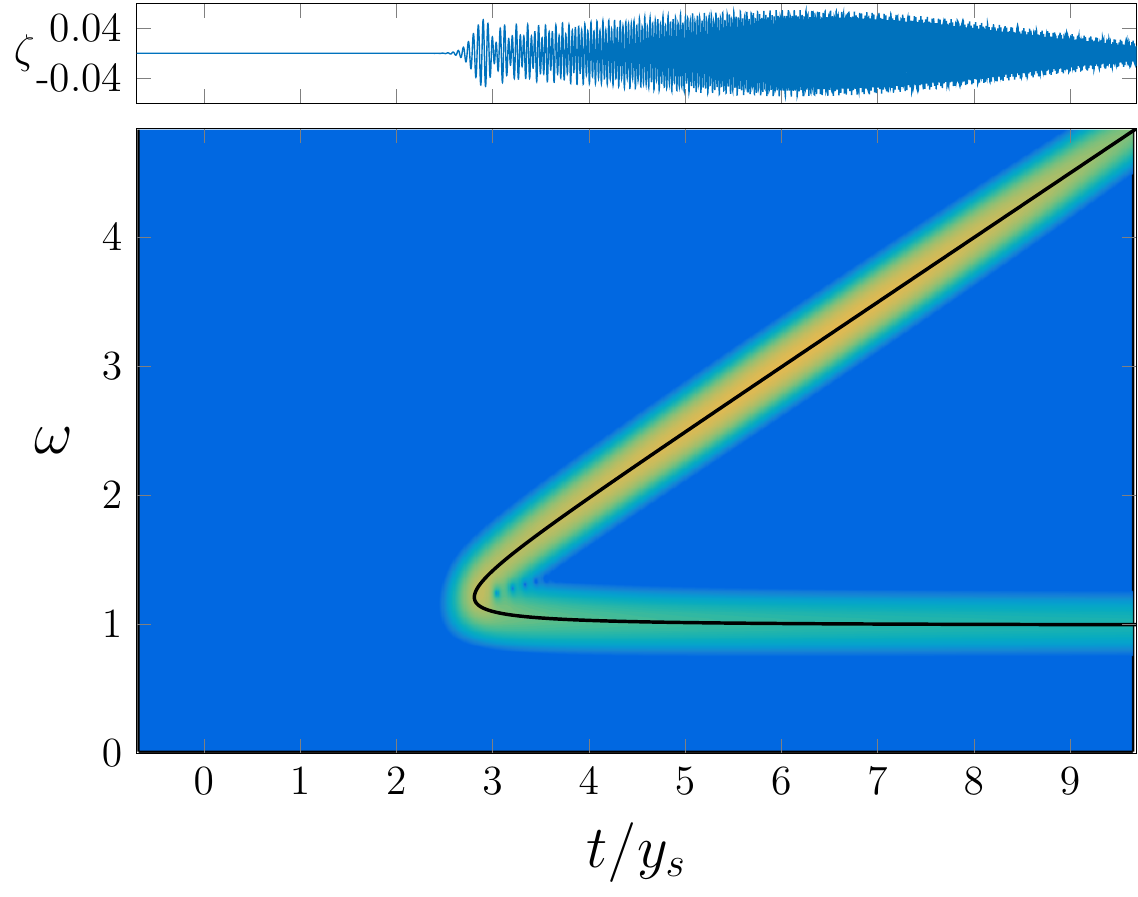}}&
\subfloat[Supercritical]{\includegraphics[width=.385\linewidth]{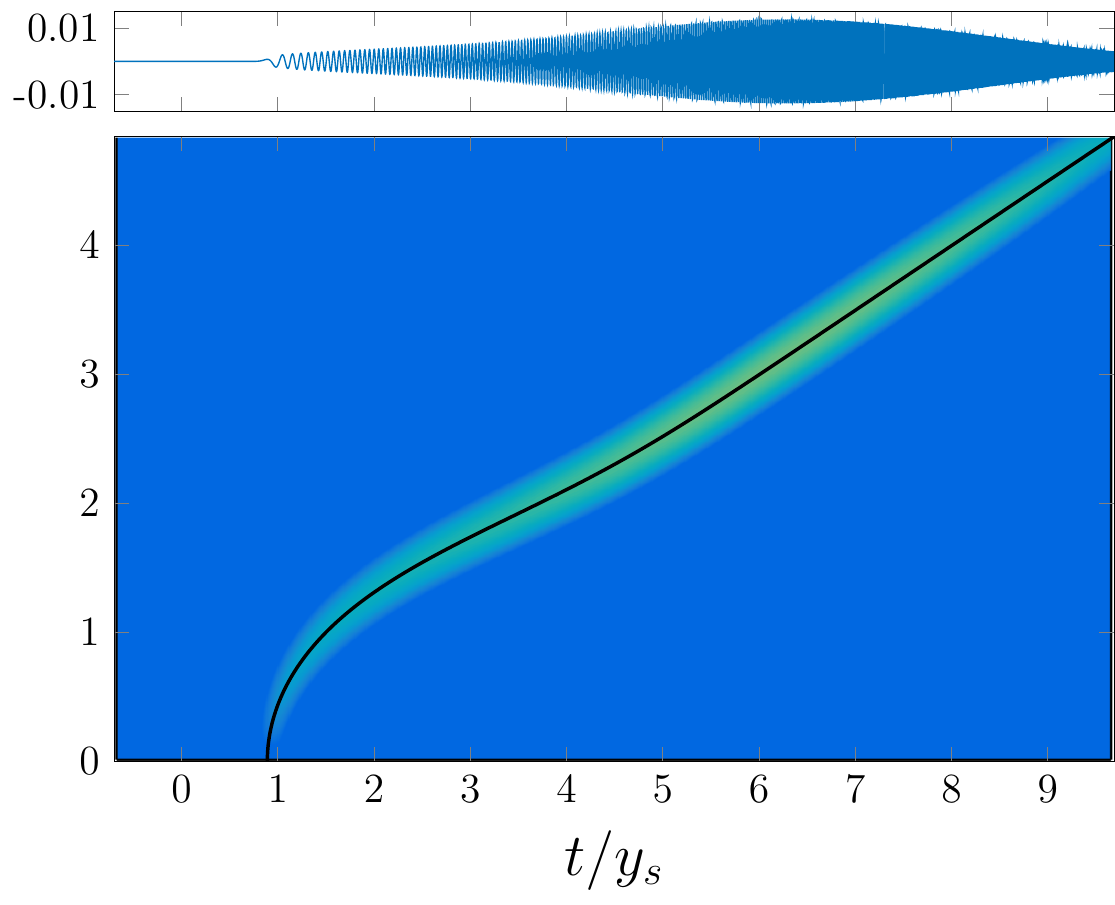}}&
\end{tabular}
\caption{Spectrograms computed for linear flow past a pressure distribution with different length based Froude numbers $F_L$ for both (a) subcritical ($F_H=0.6$) and (b) supercritical ($F_H=1.34$) flows. The black curve is the linear dispersion curve. The waves signals are presented above their respective spectrogram. {Recall, the colour intensity of the spectrograms are aligned to $\log_{10}(S(t,\omega))$.}}
\label{fig:LinSpec}
\end{figure}

We now present in Figure \ref{fig:LinSpec} spectrograms calculated from the exact solution (\ref{eq:exactLinearFinite}) to the linearised problem of flow past a pressure distribution.  In each case the cross-section of the wave pattern is defined by $y_s=100$ and the subsequent wave signal is included in the figure directly above the spectrogram.  We have chosen to present results for $F_H=0.6$, which is representative of a subcritical depth-based Froude number (far-field dimensional velocity $U$ less than $\sqrt{gH}$, where $H$ is the dimensional channel height), and $F_H=1.34$,  representative of a supercritical depth-based Froude number ($U>\sqrt{gH}$). In each regime we have shown spectrograms for four different values of the length-based Froude number $F_L$.  For a physical interpretation, recall the ``width'' of the Gaussian pressure distribution is $\delta=F_H^2/F_L^2$ so, for a fixed $F_H$, the width of the pressure decreases as $F_L$ increases.

The first observation from this figure is that in all eight cases, the high intensity region of the spectrogram appears to be centred on the linear dispersion curve described in Section~\ref{sec:dispCurve}.  Thus we can see that the linear dispersion curves do a very good job of indicating where the high intensity regions can fall.  However, we see that for low length-based Froude numbers, the high intensity regions are more prominent at lower frequencies, while for high length-based Froude numbers the high intensity regions move up to higher frequencies.  This shows that for finite-depth flows, spectrograms appear to contain more information than for infinite-depth flows, which is not unexpected given the additional parameter in the model.  This observation is of practical interest, especially if the application was to employ spectrograms of real data taken at a single point in order to unpick various features of a moving vessel.

%\afterpage{\clearpage
%\clearpage}

\section{Experimental results}\label{sec:experimental}
\begin{figure}
\centering
\includegraphics[width=\linewidth]{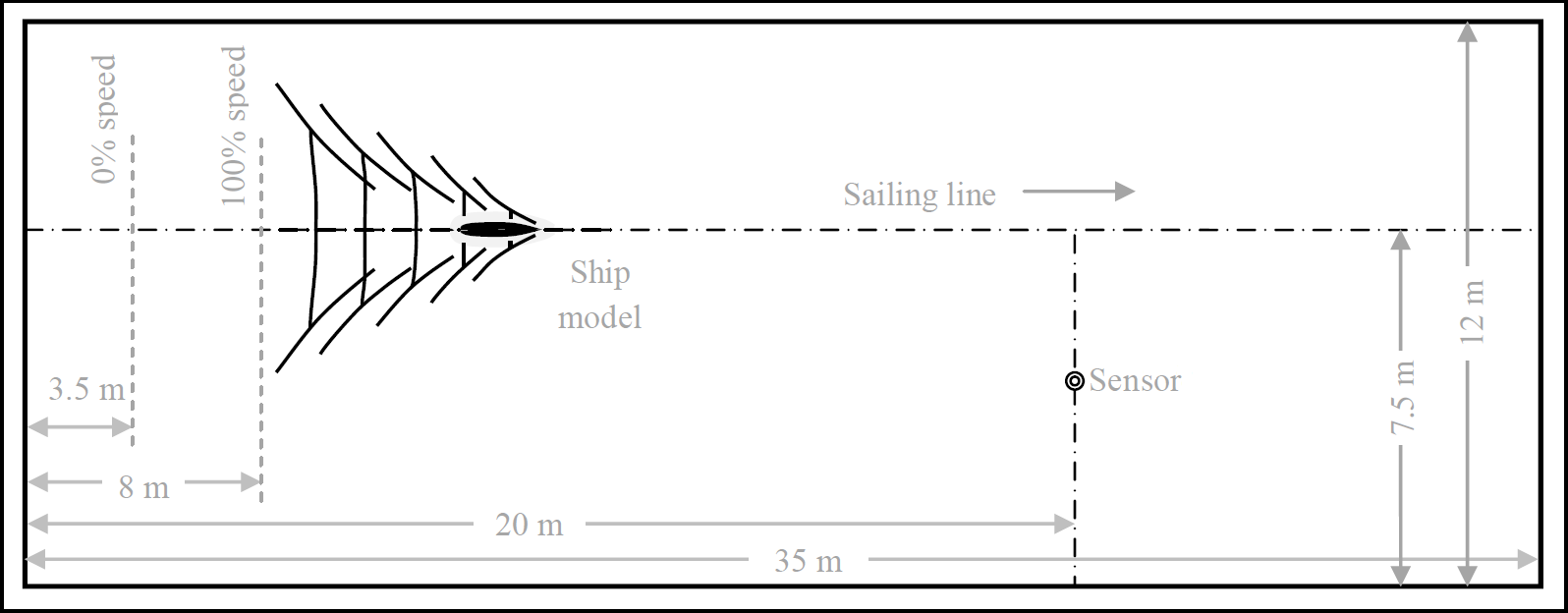}
\caption{A diagram of the experimental set up. A ship model is towed along its sailing line from rest up to a specified speed. The ship model continues to be towed at the specified speed until it passes the sensor, before decelerating. The sensor is placed 3.5 m from the sailing line.}
\label{fig:Experiment}
\end{figure}

In order to validate the linear and second-order dispersion curves against real  data, we now generate spectrograms from experiments {we} performed at the Australian Maritime College's model test basin.  The experiments were performed by towing a model ship hull down the length of a 35$\times$12~m basin (Figure \ref{fig:Experiment}). The test hulls' sailing line was offset from the centre by 1.5~m (i.e. 4.5~m on the port side and 7.5~m on the starboard side) to allow for the measurement of the ship wave to be taken at a larger distance from the ship. The sensor used in this study was located 3.5~m from the sailing line on the starboard side.  A wave-damping lane rope was placed along each of the side walls of the basin in order to reduce the amplitude of the reflected waves.  We utilise data from sixteen separate runs, four ship hulls of different aspect ratio (length divided by beam) travelling at four different speeds. The properties of the four different test hulls are provided in Table~\ref{tab:modelSpecs}.

\begin{table}
\centering
\begin{tabular}{|p{0.11\linewidth}|p{0.11\linewidth}|p{0.11\linewidth}|p{0.11\linewidth}|p{0.11\linewidth}|p{0.11\linewidth}|}
\hline
AMC model number & Length\newline ($L$, m) & Beam \newline($B$, m) & Water\newline displaced\newline ($V$, m$^3$) & Aspect\newline ratio\newline ($L/B$) & Slenderness\newline ratio\newline ($L/V^{1/3}$) \\
\hline
00-01 & 1.04 & 0.345 &$1.03\times 10^{-2}$ & 3.02 & 4.79 \\
\hline
97-10 & 1.58 & 0.305 & $1.19\times 10^{-2}$ & 5.17 & 6.91 \\
\hline
96-08 & 1.60 & 0.199 & $6.17\times 10^{-3}$ & 8.04 & 8.73 \\
\hline
99-17 & 1.83 & 0.199 & $3.81\times 10^{-3}$ & 9.18 & 11.7 \\
\hline
\end{tabular}
\caption{A table of the properties of the different experimental ship hulls. The properties are the length $L$ (m), beam $B$ (m), and the volume of water displaced at rest $V$ (m$^3$) along with the associated ratios: the aspect ratio $L/B$ and the slenderness ratio $L/V^{1/3}$.}
\label{tab:modelSpecs}
\end{table}

Spectrograms for the sixteen experimental runs are presented in Figure \ref{fig:ExperSpec}, together with the linear dispersion curves (solid curves).  We see that the linear dispersion curves do a good job of predicting the most dominant area of high colour intensity of the spectrogram.  We therefore attribute the colour intensity along the linear dispersion curve to linear ship waves generated by the hull.  This is an important observation because it provides strong evidence that the simple geometric argument in Section~\ref{sec:dispCurve} leads to an excellent ``leading order'' guess of a ship's spectrogram, regardless of the actual properties of the ship's hull.

\afterpage{\clearpage
\begin{landscape}
%\topskip0pt
%\vspace*{\fill}
\begin{figure}[h]
\centering
%\begin{tiny}
\begin{tabular}{ccccc}
\hspace{10ex}$F_H\approx0.583$&$F_H\approx1.02$&$F_H\approx1.31$&$F_H\approx2.04$&\\
\hspace{2ex}\raisebox{2ex}{\rotatebox{90}{AMC 00-01}} %$\frac{L}{B}=3.020$
\includegraphics[width=.22\linewidth]{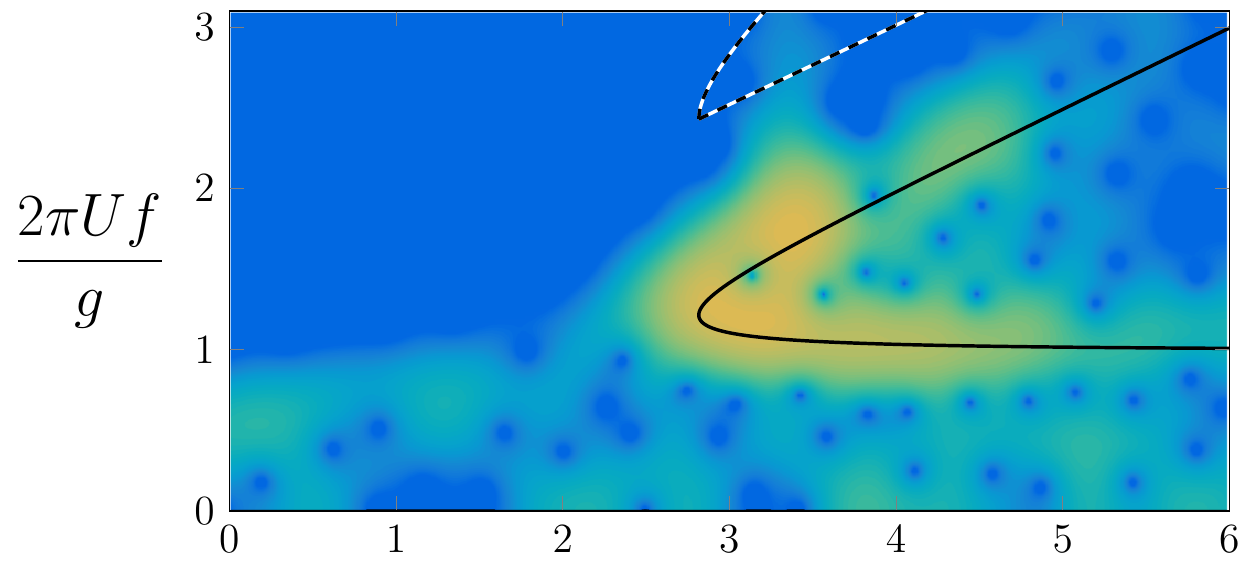}&
\includegraphics[width=.19\linewidth]{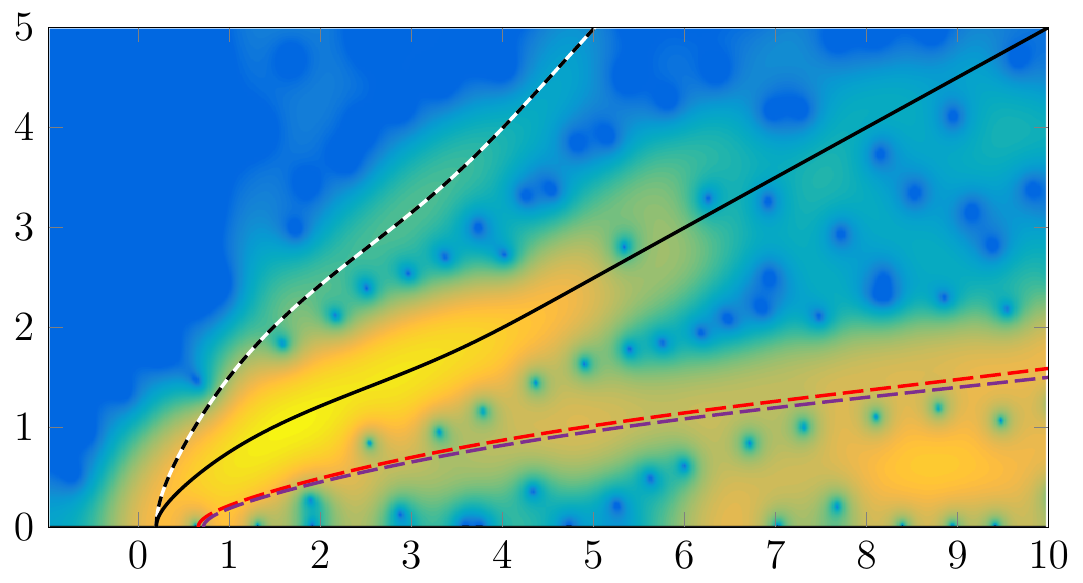}&
\includegraphics[width=.19\linewidth]{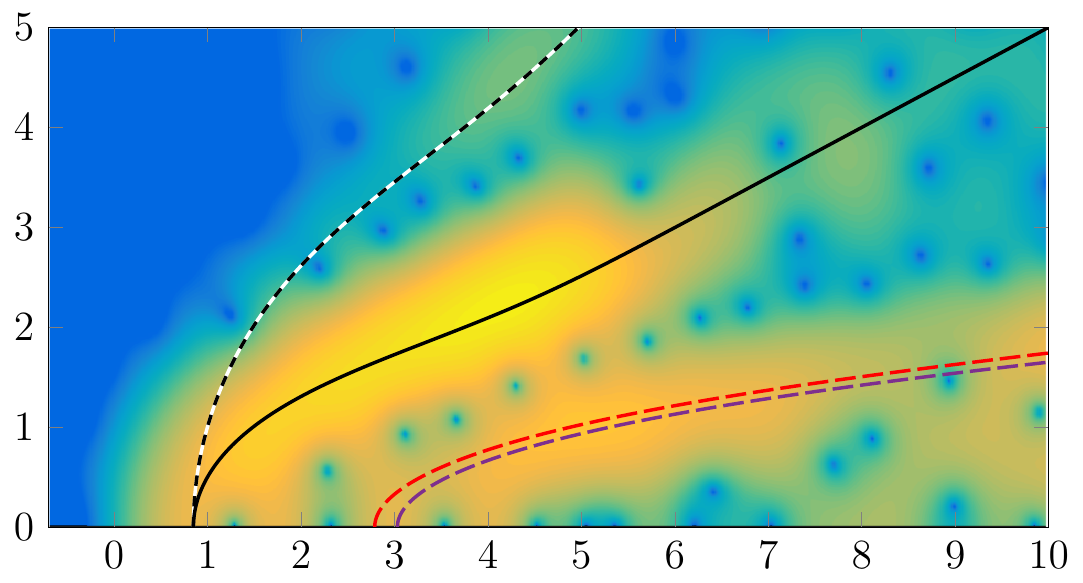}&
\includegraphics[width=.19\linewidth]{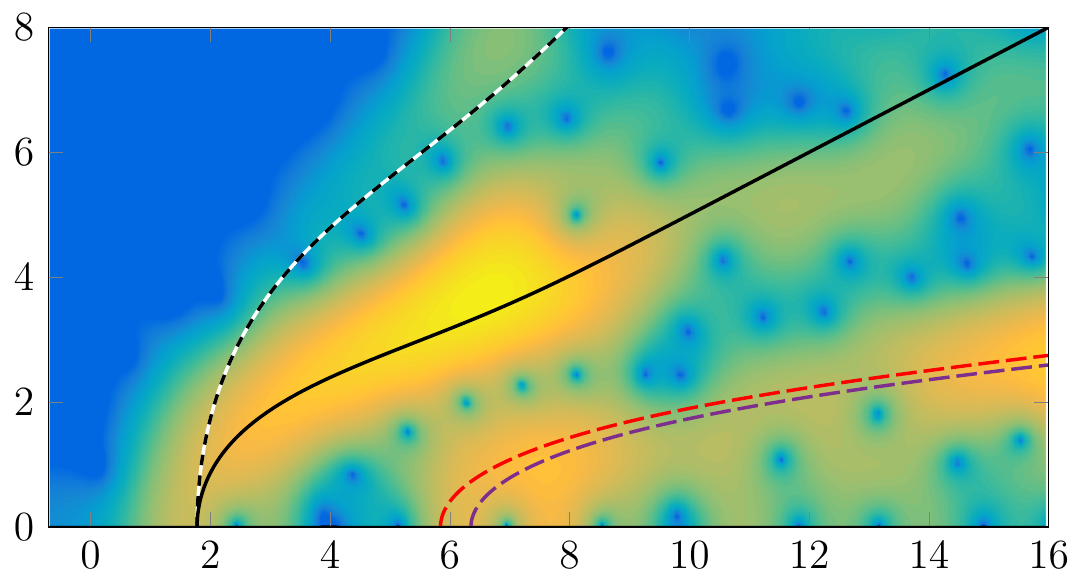}&\\
\hspace{2ex}\raisebox{2ex}{\rotatebox{90}{AMC 97-10}} %$\frac{L}{B}=5.174$
\includegraphics[width=.22\linewidth]{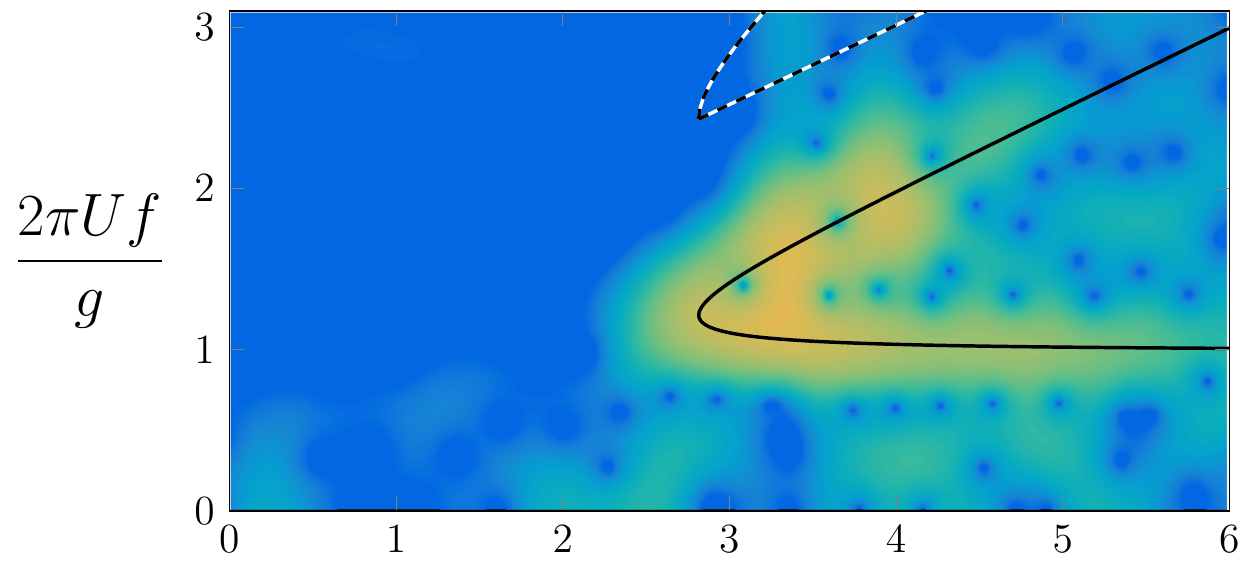}&
\includegraphics[width=.19\linewidth]{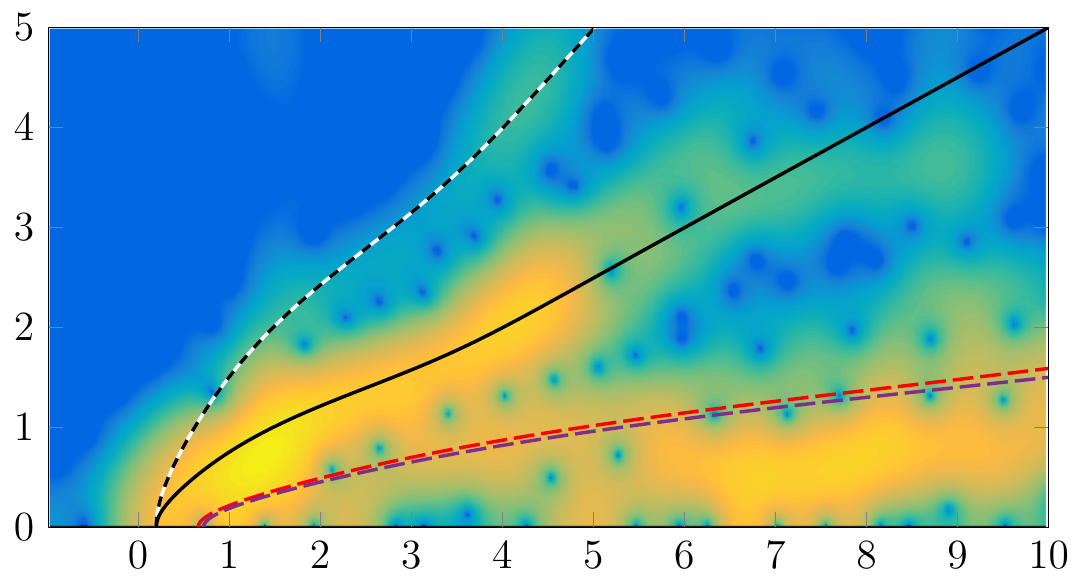}&
\includegraphics[width=.19\linewidth]{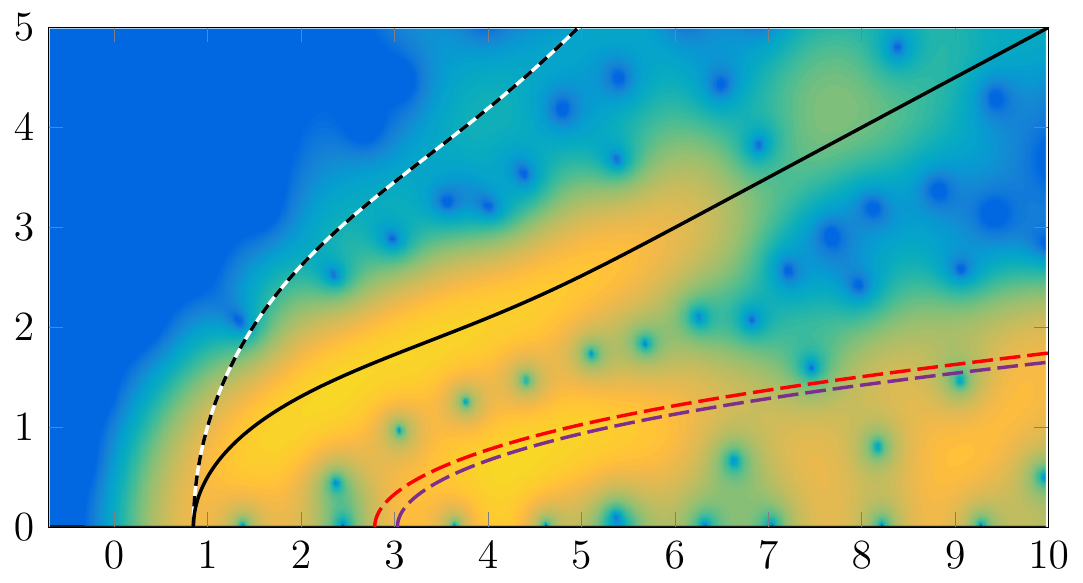}&
\includegraphics[width=.19\linewidth]{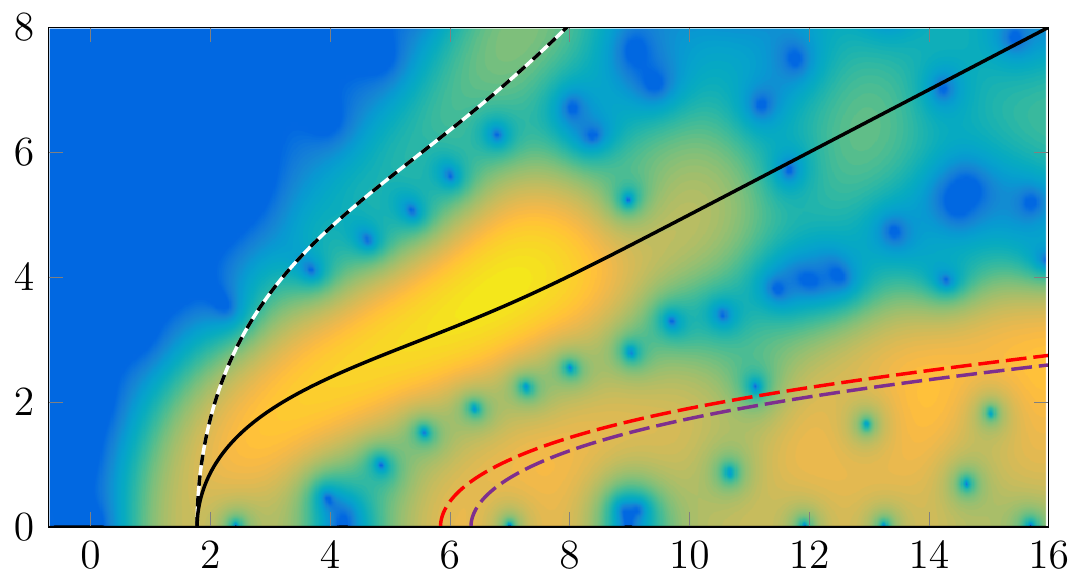}&
\hspace{-.02\linewidth}\multirow{4}{*}[0.07\linewidth]{\includegraphics[width=.03\linewidth]{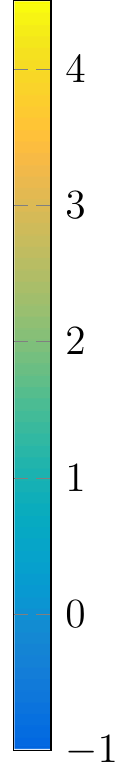}}\\
\hspace{2ex}\raisebox{2ex}{\rotatebox{90}{AMC 96-08}} %$\frac{L}{B}=8.040$
\includegraphics[width=.22\linewidth]{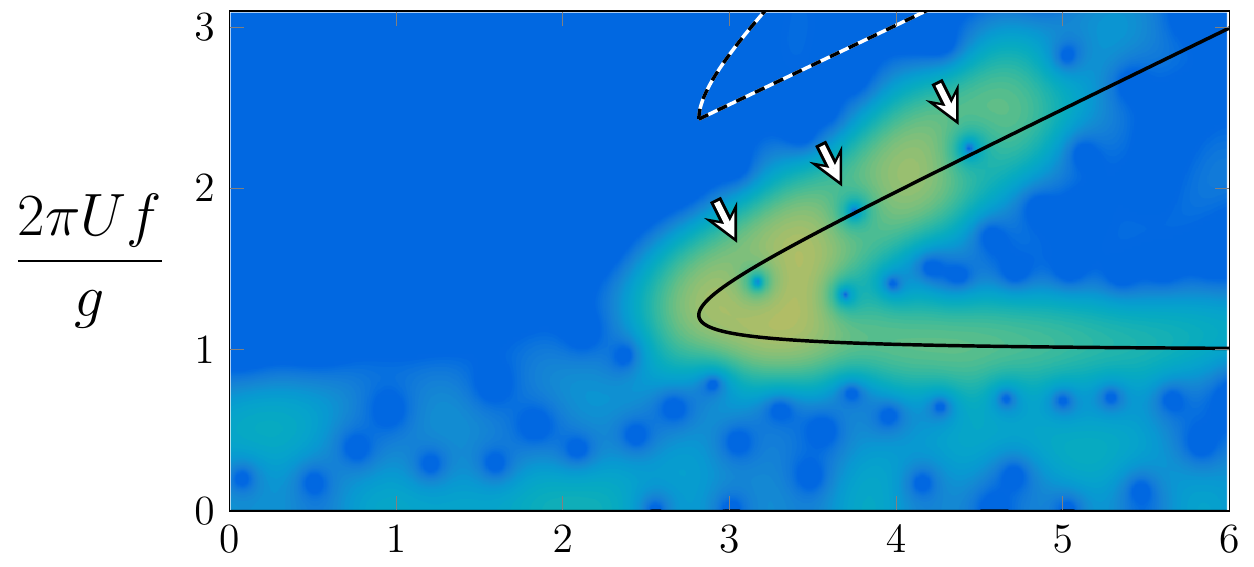}&
\includegraphics[width=.19\linewidth]{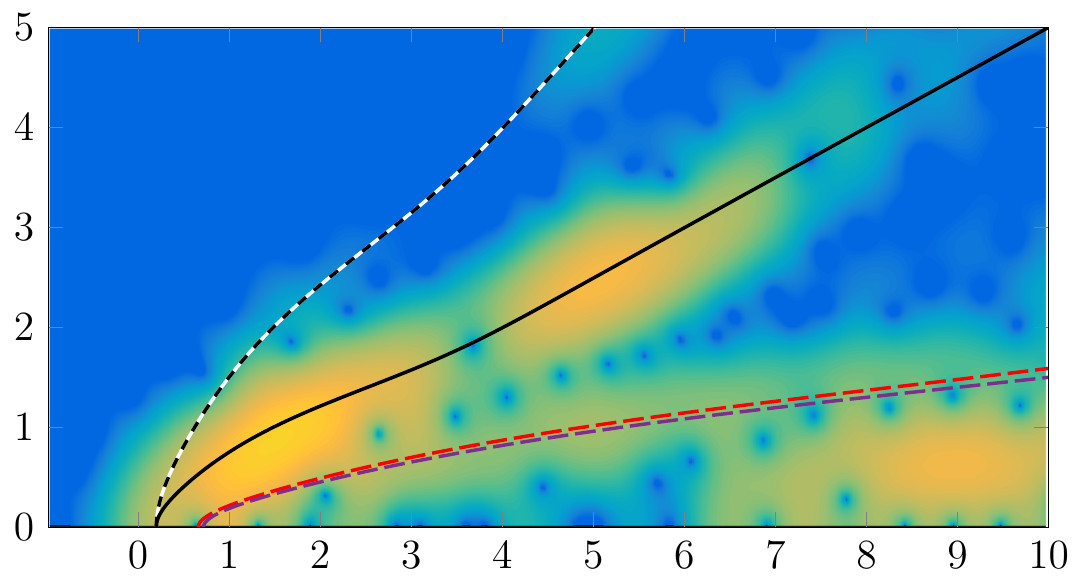}&
\includegraphics[width=.19\linewidth]{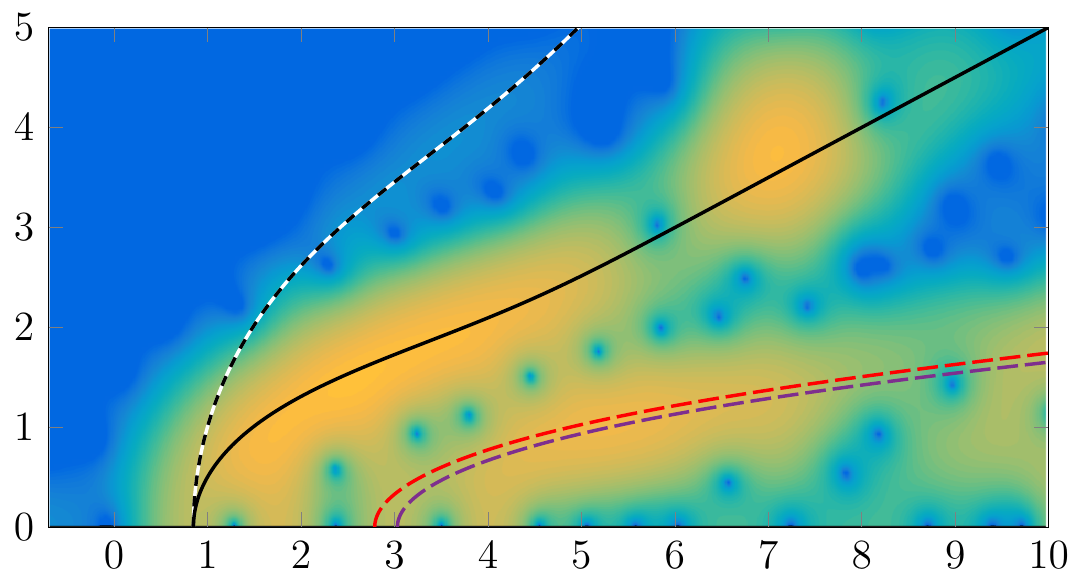}&
\includegraphics[width=.19\linewidth]{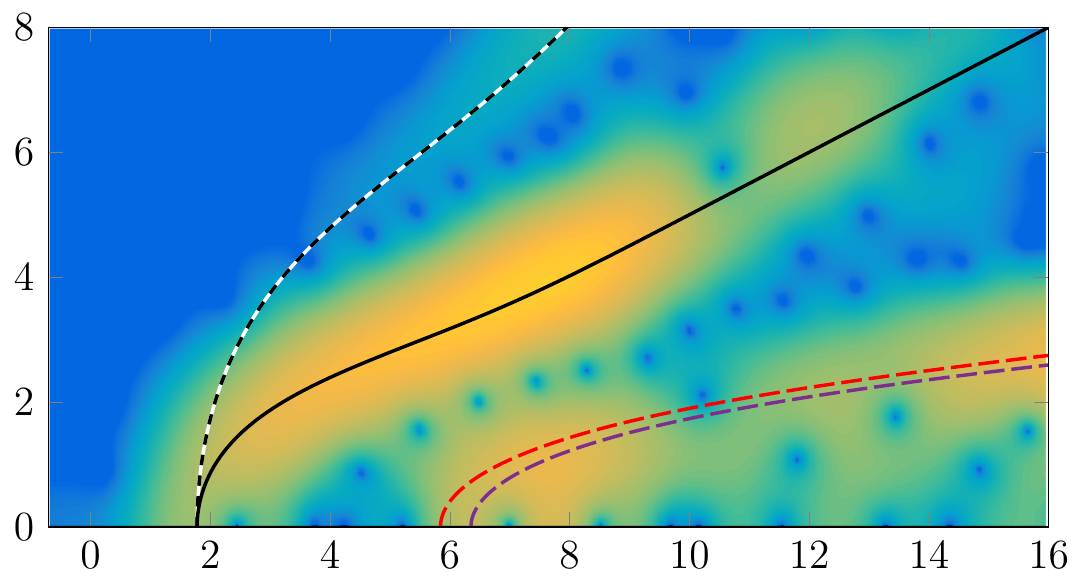}&\\
\hspace{2ex}\raisebox{4ex}{\rotatebox{90}{AMC 99-17}} %$\frac{L}{B}=9.181$
\includegraphics[width=.22\linewidth]{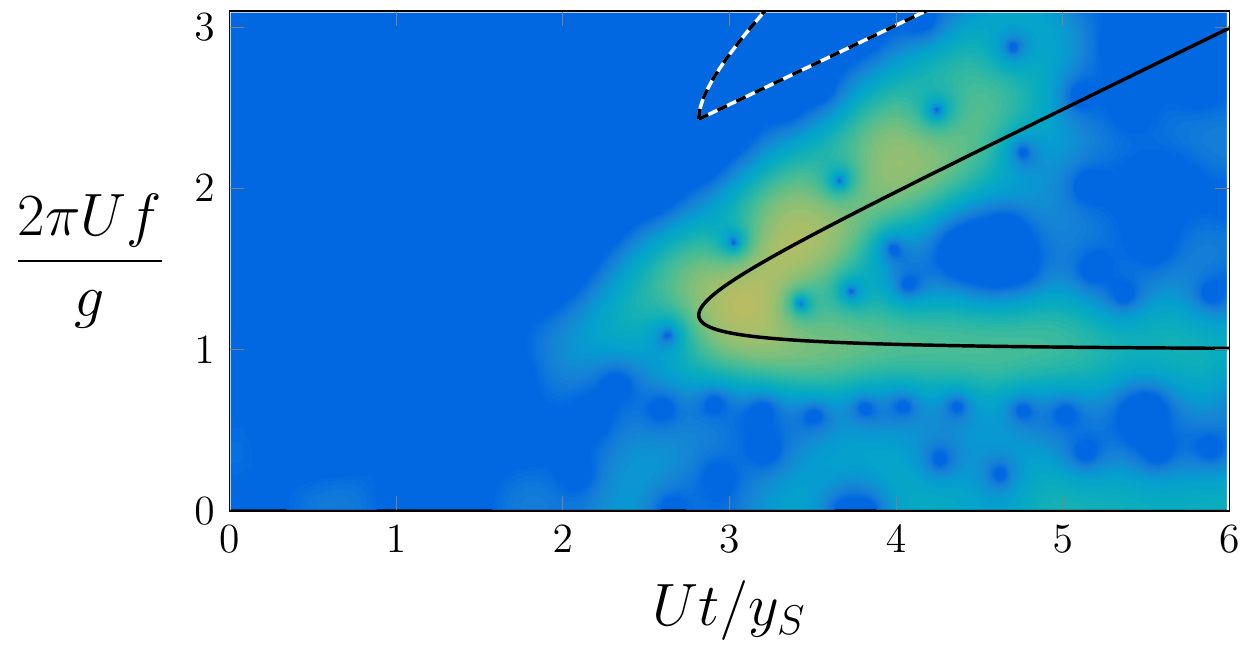}&
\includegraphics[width=.19\linewidth]{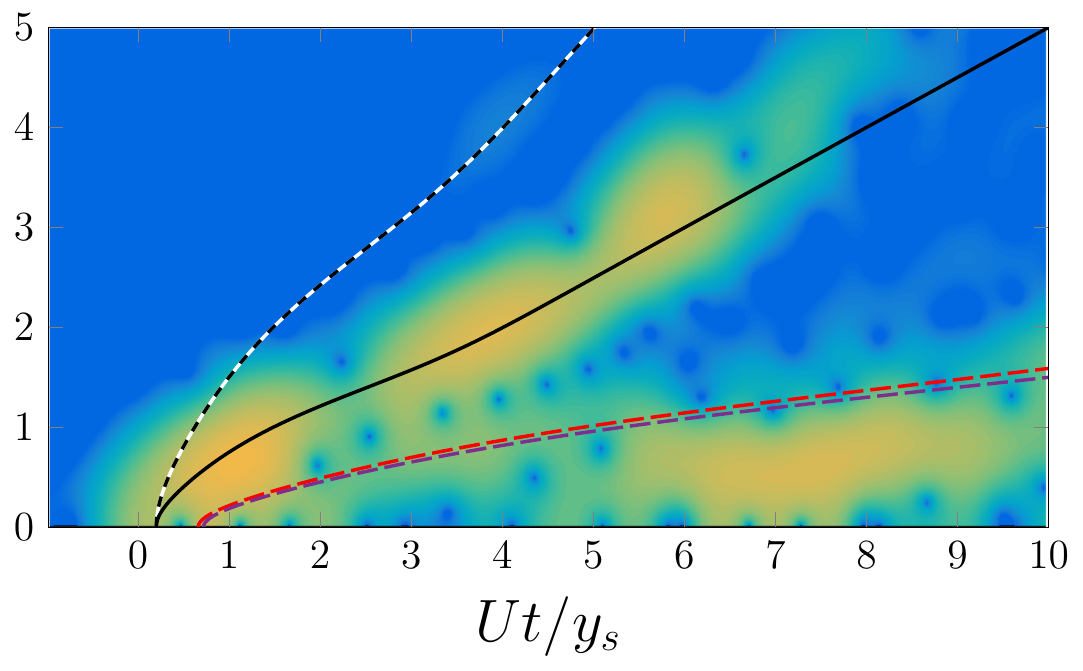}&
\includegraphics[width=.19\linewidth]{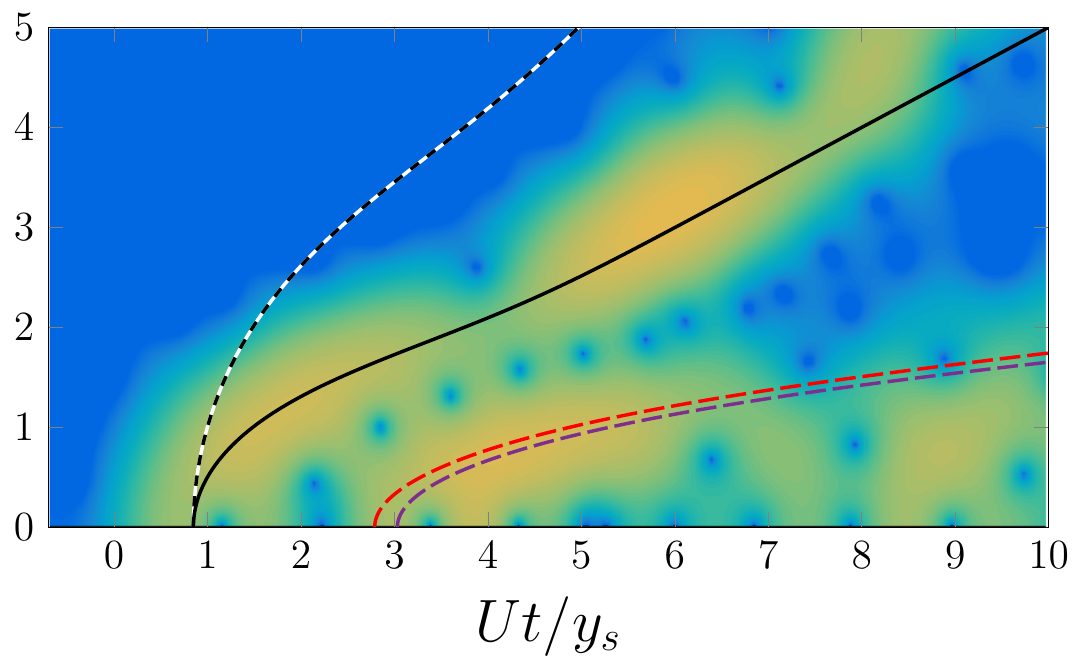}&
\includegraphics[width=.19\linewidth]{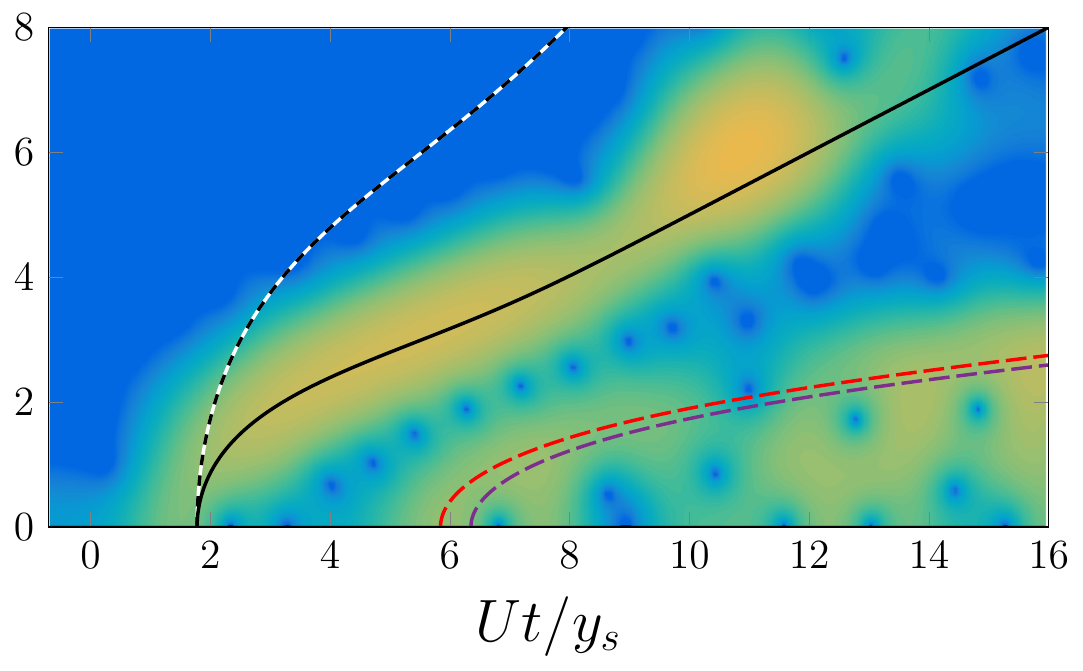}&\\
\hspace{10ex}\parbox[c][1.1em][c]{0.14\textwidth}{\centering $U=1\text{ms}^{-1}$}&
\parbox[c][1.1em][c]{0.14\textwidth}{\centering$U=1.75\text{ms}^{-1}$}&
\parbox[c][1.1em][c]{0.14\textwidth}{\centering$U=2.25\text{ms}^{-1}$}&
\parbox[c][1.1em][c]{0.14\textwidth}{\centering$U=3.5\text{ms}^{-1}$}&
\end{tabular}
%\end{tiny}
\caption{Spectrograms computed from experimental data for the ship hulls given in Table \ref{tab:modelSpecs}, and speed $U$. The experimental hulls (rows) are in ascending order of aspect and slenderness ratio. The measurements were taken a distance $y_s=3.5$~m from the sailing line. The black curves are the linear dispersion curves (\ref{eq:observedFrequency}).  The {black and white} dashed curves are the second-order dispersion curves $\omega_{11+}$ and $\omega_{12+}$ for the first column (subcritical flow) and $\omega_{11+}$ for the second, third and fourth columns (supercritical flow).  The long-dashed curves are the dispersion curves for the wake reflected off the starboard {(upper curve)} and port {(lower curve)} side walls of the model test basin. Note that the axes labels are given as dimensionless angular frequency $\omega=2\pi Uf/g$ and scaled time $Ut/y_s$ where frequency $f$, time $t$ and distance $y_s$ in the axis are \emph{dimensional} quantities. {Recall, the colour intensity of the spectrograms are aligned to $\log_{10}(S(t,\omega))$.}}
\label{fig:ExperSpec}
\end{figure}
%\vspace*{\fill}
\end{landscape}
\clearpage}

We now consider nonlinear effects on the spectrograms by including the second-order dispersion curves ({ black and white} dashed curves) in Figure~\ref{fig:ExperSpec}.  For subcritical flows ($F_H<1$), we only show the second-order dispersion curves $\omega_{11+}$ and $\omega_{12+}$ because they were the most prevalent in the nonlinear numerical simulations performed by Pethiyagoda et al.~\cite{pethiyagoda17} for the infinite-depth counterpart.  For supercritical flows ($F_H>1$) there is only one second-order curve.  The first observation about these second-order curves is that in the subcritical examples they do not coincide with high intensity regions of the spectrograms, which suggests that the four examples we present in this regime are essentially linear flows.  Turning to supercritical flows, we see the second-order dispersion curves align with areas of colour intensity in the spectrogram, especially for lower aspect ratio hulls (AMC 00-01 and AMC 97-10).  Thus we propose that the examples for AMC 00-01 and AMC 97-10 have higher amplitude waves which are weakly nonlinear in nature.  Indeed, it seems that nonlinearity is strongest for the lower aspect ratio hulls we use in our experiments.  Further, we postulate that ship wave patterns are more nonlinear if the volume of fluid displaced by the surface-piercing hull is higher.  This notion supports the modelling idea that the strength of an applied pressure distribution can act as a proxy for the volume of water displaced by a hull.

For the supercritical flows ($F_H>1$), there is a significant amount of colour intensity occurring away from the linear and second-order dispersion curves that cannot be explained by our linear or weakly nonlinear theory.  Instead, we believe much of this intensity is due to wave reflection.  To support this idea, we have plotted two additional linear dispersion curves (long-dashed curves) representing ship waves that have reflected perfectly off the starboard and port side walls of the model test basin.  Thus, in essence we assume the existence of two ``ghost'' ships travelling in line with the experimental ship, but further away from the sensor. The reflected curves are computed by recalling that the shape of the dispersion curve depends on the depth-based Froude number and is a function of $t/y_s$, where $y_s$ is the minimum distance to the sailing line. Given that each additional ghost ship has the same depth based Froude number, the reflected dispersion curve can be expressed by rescaling the horizontal axis by $y_s/y_r$ (that is if $\omega(t/y_s)$ is the linear dispersion curve, $\omega(t/y_r)$ is the reflected dispersion curve), where $y_r$ is the distance from the sensor to the ghost ship's sailing line.  We see from Figure~\ref{fig:ExperSpec} that the reflected dispersion curves seem to align with much of the colour intensity of the larger Froude number spectrograms, providing strong evidence that these regions of high intensity are in fact due to wave reflection.  This alignment is quite promising given the assumption of perfect reflection, which may be too idealised to accurately reflect the experimental spectrograms (the wave-damping ropes on the walls of the basin are likely to decrease the amplitude of the reflected waves but only have a minor effect on their wavelength).  We note that there is no evidence of wave reflection in the subcritical examples ($F_H<1$) because the reflected waves had not reached the sensor by the time the sensor had finished recording the surface height data.

{
Finally, we note an additional region of high colour intensity at low frequency and large time for the supercritical experiments (in particular, see the results for $F_H\approx 1.02$). One potential explanation for this additional region is a secondary reflection wave (this is, the reflected wave being reflected again off of the other wall). However, under the current assumption of perfect reflection, the secondary reflection curves do not appear to align with this additional region of intensity (not shown) and instead a more realistic model of reflection may be appropriate. Another possible explanation could relate to the waves caused by the slow vertical motion of the wave-damping lane ropes. These motions have been observed being induced by leading longer period ship waves, generally for low aspect ratio hulls. The effect of non perfect reflection or motion of the wave-damping lane ropes on a spectrogram is presently unknown and is the subject of future work.
}

%\FloatBarrier

\section{Wave interference effects}\label{sec:interference}

There is one more feature of our experimental spectrograms that we wish to investigate further.  We see in the subcritical examples in Figure~\ref{fig:ExperSpec} that there appears to be periodic regions of low intensity (blue dots) that roughly follow the sliding-frequency ($\omega_1$) mode.  For example, the spectrogram for AMC 96-08 with $F_H\approx 0.583$ has three white arrows pointing to such dots.  A  potential explanation for these oscillations in colour intensity is the wave interference that is well known to occur from the interaction between waves generated by the bow and stern of a ship's hull \cite{he14,ma16,miao15,noblesse14,zhang15a,zhu15}.

To observe the effects of wave interference in our mathematical model, we consider a two-point wave-maker which consists of two Gaussian pressure distributions, one ahead of the other.  An exact solution for linear ship waves caused by such a two-point wave-maker is given by
\begin{equation}
Z(x,y)=\zeta\left(x+\frac{\ell}{2},y\right)-\zeta\left(x-\frac{\ell}{2},y\right)\label{eq:twoPointWavemaker}
\end{equation}
where $\zeta(x,y)$ is given by equation (\ref{eq:exactLinearFinite}) and $\ell$ is the nondimensional distance between the two pressure distributions.
\begin{figure}[htb]
\centering
\raisebox{2ex}{\includegraphics[width=.4\linewidth]{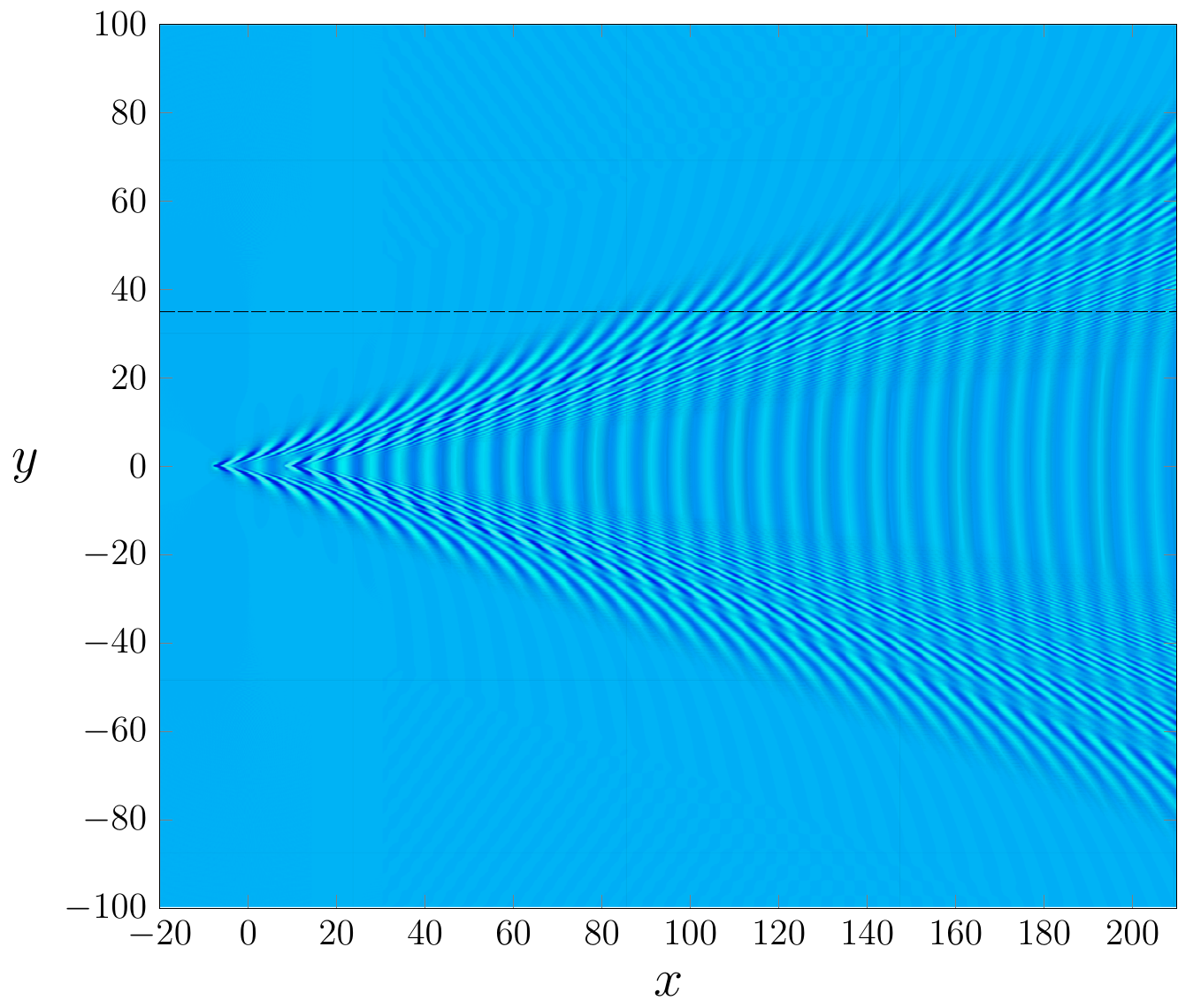}}\hfil
\includegraphics[width=.46\linewidth]{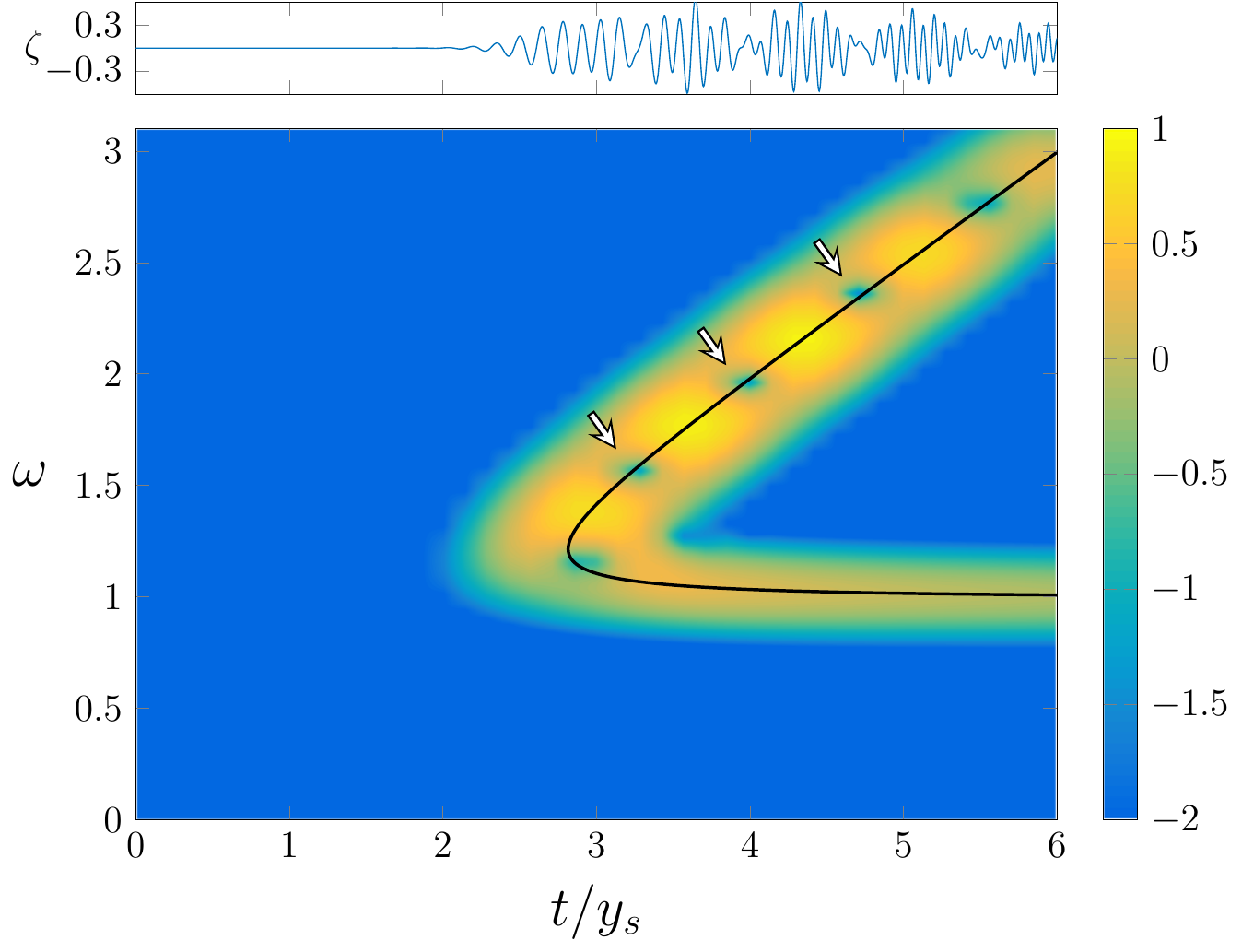}
\caption{A plan view of a free-surface profile for a ship wave computed using a two-point wave-maker (\ref{eq:twoPointWavemaker}) and its associated spectrogram, where $F_H=0.6$, $\ell=16$, $F_L=1$, $\delta=0.36$ and $y_s=35$, which approximate the values for the AMC 98-08 hull travelling at 1 ms$^{-1}$. The dashed line indicates the location of the cross-section taken to form the signal used for the spectrogram. The signal is presented above the spectrogram.  The white arrows on the spectrogram point to periodic regions of low intensity (blue dots). {Recall, the colour intensity of the spectrogram is aligned to $\log_{10}(S(t,\omega))$.}}
\label{fig:Lin2point}
\end{figure}
In Figure \ref{fig:Lin2point} we provide an example of a spectrogram generated from the wake created by a two-point wave-maker with $F_H=0.6$, $F_L=1$, $\ell=16$ and $\delta=0.36$ (these are approximate parameter values for the AMC 98-08 hull travelling at 1 ms$^{-1}$).  We can clearly see the almost-periodic nature of the colour intensity (including blue dots that are indicated by the white arrows) along the sliding frequency mode that occurs as a result of including a second pressure distribution.  Further, we can match the oscillating low and high intensity regions with the wave signal itself, which is placed directly above in the figure.

The results in Figure \ref{fig:Lin2point} are for one example only, but we can infer from our mathematical model that wave interference is likely to manifest itself in a spectrogram for a real vessel with a large aspect ratio.  This is consistent with our observations of experimental results in Figure~\ref{fig:ExperSpec}, since the periodic blue dots along the sliding frequency mode appear more prominent as the aspect ratio increases.

\section{Discussion}

Broadly speaking, Fourier analysis has recently been adopted as a tool for understanding the underlying features of a measured ship wake \cite{caplier16,gomit14,didenkulova13,sheremet13,torsvik15a,torsvik15b}.  One form of analysis is performing a two-dimensional Fourier transform on the measured surface height~\cite{caplier16,gomit14}. While this approach can provide information for the entire ship wave (both near and far field), the methods for measuring the free-surface height (e.g. the use of fluid seeded particles \cite{chatellier13} or a projected laser grid \cite{gomit15}) can be impractical in real world settings. In contrast, in the present study we are interested in a form of time--frequency analysis that involves performing many short-time Fourier transforms on a cross-section of a ship's wake and plotting the result as a spectrogram \cite{brown89,didenkulova13,sheremet13,torsvik15b,torsvik15a,wyatt88}.  The key advantage of this approach is that the required data can be collected at a single fixed sensor, which in practice is a relatively simple exercise.

We have considered a mathematical model for generating a ship wake where the ship itself is represented by a pressure distribution applied to the surface of the water.  We have used geometric arguments and an exact solution (of the linearised problem) to predict where the dominant regions of high intensity will fall on a spectrogram from this model.  By varying the depth-based Froude number, we have shown how transverse and divergent waves appear on spectrograms for both subcritical and supercritical regimes.  The effects of nonlinearity, wave reflection and wave interference have also been considered.  By comparing with experimental results {we have generated} for realistic hull shapes, we have provided a variety of evidence that shows how our mathematical model together with spectrogram analysis has the potential to explain various properties of real vessels.

In terms of future work, there is a need to explore in some detail the effect of using a more realistic mathematical model that takes into account specific design features of realistic ship hulls.  This includes further studying the effect of wave interference for long and thin ship hulls and vessels with multiple hulls.  The issue of nonlinearity needs further attention, especially in terms of solitary waves (precursor solitons) that propagate ahead of a vessel moving at near-critical speed ($U\approx \sqrt{gH}$) \cite{soomere07,torsvik15a}.  From an experimental perspective, it would be helpful to compare theoretical spectrograms with data measured on a much wider experimental facility, so as to eliminate the effects of wave reflection from the bounding walls. { Alternatively, identifying the spectrogram features caused by the model test basin (ie. wave reflection or slow motion of the wave-dampening lane ropes) will allow for the isolation of the true ship waves.}  Finally, there is scope to expand upon this work to develop a framework in which surface height data taken from a sensor (or a set or sensors) can be used to identify not only the position of a ship as it passes by, but also various properties of the ship's hull (such as the length, beam, water displaced, aspect ratio, and so on).

\section*{Acknowledgement}
\noindent SWM acknowledges the support of the Australian Research Council via the Discovery Project DP140100933. GJM acknowledges the support of the Australian Research Council via the Linkage Project LP150100502. RP acknowledges the support of a Lift-off Fellowship from the Australian Mathematical Society.

\section*{References}
\bibliographystyle{plain}
%\bibliography{references}

\end{document}